\documentclass[sn-nature]{sn-jnl}

\usepackage[vietnamese, english]{babel}
\usepackage{braket}

\usepackage{graphicx}%
\usepackage{multirow}%
\usepackage{amsmath,amssymb,amsfonts}%
\usepackage{amsthm}%
\usepackage{mathrsfs}%
\usepackage[title]{appendix}%
\usepackage{xcolor}%
\usepackage{textcomp}%
\usepackage{manyfoot}%
\usepackage{booktabs}%
\usepackage{algorithm}%
\usepackage{algorithmicx}%
\usepackage{algpseudocode}%
\usepackage{listings}%

\usepackage{bm}

\theoremstyle{thmstyleone}%
\theoremstyle{thmstyletwo}%

\theoremstyle{thmstylethree}%

\begin{document}

\title[Topological bound states in the Haldane model]{Emerging topological bound states in Haldane model zigzag nanoribbons}


\author*[1]{\fnm{Simone} \sur{Traverso}}\email{simone.traverso@edu.unige.it}

\author[1,2]{\fnm{Maura} \sur{Sassetti}}\email{sassetti@fisica.unige.it}

\author[1,2]{\fnm{Niccol\`o Traverso} \sur{Ziani}}\email{traversoziani@fisica.unige.it}

\affil*[1]{\orgdiv{Physics Department}, \orgname{University of Genoa}, \orgaddress{\street{Via Dodecaneso 33}, \city{Genova}, \postcode{16146}, \country{Italia}}}

\affil[2]{\orgname{CNR-SPIN}, \orgaddress{\street{Via Dodecaneso 33}, \city{Genova}, \postcode{16146}, \country{Italia}}}


\abstract{Zigzag nanoribbons hosting the Haldane Chern insulator model are considered. In this context, a reentrant topological phase, characterized by the emergence of quasi zero dimensional in-gap states, is discussed. The bound states, which reside in the gap opened by the hybridization of the counter-propagating edge modes of the Haldane phase, are localized at the ends of the strip and are found to be robust against on-site disorder. These findings are supported by the behavior of the Zak phase over the parameter space, which exhibits jumps of $\pi$ in correspondence to the phase transitions between the trivial and the non-trivial phases. The effective mass inversion leading to the jumps in the Zak phase is interpreted in a low energy framework. Setups with non-uniform parameters also show topological bound states via the Jackiw-Rebbi mechanism. All the properties reported are shown to be extremely sensitive to the strip width.}




\maketitle

\section{Introduction}\label{intro}
The discovery of topological insulators and topological superconductors completely revolutionized the usual classification of the phases of matter, shedding light on the fact that the Ginzburg-Landau classification was but a partial description~\cite{Zhang_rev_2011}. Starting from the first feasible proposal for a topological insulator \cite{bernevig_2007}, in less than two decades the field has undergone several major breakthroughs, like the classification of topological phases by symmetry classes \cite{ryu_2010}, the discovery of higher order topology~\cite{Neupert_2018} and, very recently, of non-Hermitian topology~\cite{Bergholtz_2021}. {\color{black} In this context, the dimensionality of the systems plays a crucial role for the definition of the topological invariants. Indeed, these are typically defined relative to systems compactified in all directions. The bulk boundary correspondence, then, suggests that when the system is non-trivial, and is made semi-infinite in one direction, metallic states appear at the boundary. These metallic states persist as long as the uncompactified dimension of the system does not become comparable with their decay length.

The improvement in the nanostructuration of topological phases of matter has made it possible to realize samples in which this condition is not met~\cite{Molenkamp_2020, Claessen_2022}. In this regard,} a very active branch of research is nowadays related to the study of finite size effects on topological phases~\cite{niu_2008, PhysRevB.92.235407, PhysRevB.98.205129, Claessen_2022, Teemu_2011, Molenkamp_2020, Rodriguez2020, Fleckenstein2021, Traverso_2022, Vigliotti_2023, PhysRevB.107.245409, Zhu_2023, Saha_2023}. Indeed, dimensional crossovers between topological phases is a promising way to engineer novel topological systems~\cite{Potter_2010, Zhang2022}. In this context, the way has been paved by the extensive studies performed on graphene nanoribbons (GNRs), that have revealed an extremely rich phenomenology. For instance, it was found that these systems, depending on the nanoribbon width and on the nature of its terminations, can host robust topological bound states~\cite{PhysRevLett.119.076401, Li_2021} amenable to detection with local probes~\cite{local1,local2,local3}.

Taking a step back, even before the first theoretical proposal for a symmetry protected topological insulator was conceived, many topologically non-trivial systems had been object of study, first of all the quantum Hall system~\cite{Klitzing_1980}.
In that context, a milestone for the comprehension of topological phases had been conceived: the Haldane model~\cite{haldane_1988}. Dated back to 1988, it represented the first theoretical proposal for a system realising a quantum anomalous Hall phase and is now regarded as the most famous model for a Chern insulator. Even more importantly, the time reversal doubling of the Haldane model results in the Kane-Mele model which, describing spinful fermions on a honeycomb lattice with strong spin-orbit interaction, is the prototypical model of a time reversal protected topological phase~\cite{Kane-Mele-1_2005, Kane-Mele-2_2005}. Although the Kane-Mele model was originally proposed to describe the electrons in graphene, where the predicted spin Hall phase cannot be observed since the spin-orbit coupling is too small to yield a sizable topological gap, it has recently found a direct experimental realization in Bismuthene~\cite{Reis_2017} {\color{black} and Germanene~\cite{Bampoulis_2023}}, and is hence receiving renewed attention.

In light of this, Haldane model nanoribbons represent a significant model for studying the physics of these newly discovered honeycomb-based topological materials. Even more importantly, they configure themselves as the optimal theoretical playground for merging the physics of nanostructured topological insulators and the one inherited from graphene nanoribbons. In this paper, we focus on zigzag Haldane nanoribbons and assess the effects of dimensional reduction on the topological phase of the Haldane model. We find that, for thin enough strips, there are multiple regions of the parameter space in which the chiral edge states gap out and, correspondingly, degenerate pairs of quasi zero-dimensional (0D) end-states appear whose energy lies inside the gap. Such regions however are intercalated, through topological quantum phase transitions, to phases without bound states. We hence unveil a complex, width-dependent, reentrant quantum phase diagram, which we characterize by numerically computing a well established indicator for the classification of topological phases in (effectively) one-dimensional systems, that is, the Zak phase~\cite{Zak_1989, Jens_2017,Jeong2022}. Moreover, {\color{black} we explain the mechanism leading to the mass inversion via a phenomenological low energy theory, effectively modelling the chiral edge states coupling. Finally,} we show that domain walls in the on-site staggered potential distribution can localize quasi 0D bound states. These Jackiw-Rebbi like states~\cite{Jackiw_1976}, ubiquitous in topological phases of matter~\cite{Qi_2008, Teemu_2011, Ziani_2017, Fu2022, Cheng2023}, retain a fractional charge of $\pm \frac{e}{2}$ ($e$ the electron charge) and are found to be robust against the introduction of random on-site disorder. More generally, our results demonstrate that the physics of coupled topological edges can be way richer than what is expected from naive low energy theories.

\section{Results}
\label{sec:Res_disc}
\subsection{Haldane model on zigzag nanoribbons}
The Haldane model~\cite{haldane_1988} describes spinless fermions on a honeycomb lattice, pierced by an orthogonal periodic magnetic field and having the full symmetry of the lattice and zero net flux over the unit cell. Time reversal symmetry is broken for the model, so that the transverse Hall conductance can be non-trivially quantized~\cite{bernevig2013topological}. We denote the two sublattices of the honeycomb lattice as $A$ and $B$ and we choose as primitive vectors $\mathbf{a}_1 = a (1,0) $ and $\mathbf{a}_2 = a(\frac{1}{2},\frac{\sqrt{3}}{2})$. Moreover, we place the unit cell origin on the $A$ sites, so that the basis vectors are given by ${\bm \delta}_a = (0,0)$ and ${\bm \delta}_b = a(\frac{1}{2},\frac{1}{2\sqrt{3}})$. The Hamiltonian is thus given by
\begin{equation}
    \begin{split}
        \mathcal{H} &= t_1 (\sum_{l,n} a^\dagger_{ln} b_{ln} + a^\dagger_{ln} b_{l-1n}+a^\dagger_{ln}b_{ln-1})+\text{h.c.}\\
        &+t_2 \mathrm{e}^{\mathrm{i}\phi} (\sum_{l,n} a_{ln}^\dagger a_{l+1n} + a^\dagger_{ln}a_{l-1n+1}+a^\dagger_{ln}a_{ln-1})+\text{h.c.}\\
        &+t_2 \mathrm{e}^{-\mathrm{i}\phi} (\sum_{l,n} b_{ln}^\dagger b_{l+1n} + b^\dagger_{ln}b_{l-1n+1}+b^\dagger_{ln}b_{ln-1})+\text{h.c.}\\
        &+m (\sum_{l,n} a^\dagger_{ln}a_{ln}-b^\dagger_{ln}b_{ln}),
    \end{split}
\end{equation}
where $a_{ln}$ ($b_{ln}$) destroys a fermion at $l\mathbf{a}_1+n\mathbf{a}_2+\bm{\delta}_{a(b)}$; $t_1$ and $t_2$ parameterize the intensity of the nearest and next to nearest neighbor hoppings respectively; $\phi$ is a phase accounting for the staggered magnetic flux inside the unit cell and $m$ tunes a term of staggered on-site potential breaking inversion symmetry. In the topological phase, that occurs for $|m/t_2|<3\sqrt{3}\sin\phi$, the bulk bands are gapped and the Chern number~\cite{bernevig2013topological} has a non-trivial value ($c=\pm 1)$~\cite{haldane_1988}.

Correspondingly, in a strip geometry chiral modes localized on opposite edges and with gapless dispersion relation occur, in accordance with the bulk boundary correspondence. This can be explicitly seen by imposing periodic boundary conditions (PBC) in the $\mathbf{a}_1$ direction and going in $k$-space. For a strip of length $L=Na$, the Fourier transformation is defined as
\begin{align}
    a_{ln}&= \dfrac{1}{\sqrt{N}}\sum_{k} \mathrm{e}^{-\mathrm{i}kx_{ln}^a}a_n(k) & x_{ln}^a&=la+\frac{n}{2}a, \\
    b_{ln}&= \dfrac{1}{\sqrt{N}}\sum_{k} \mathrm{e}^{-\mathrm{i}kx_{ln}^b}b_n(k) & x_{ln}^b&=la+\frac{n+1}{2}a,
\end{align}
with the Bloch momenta discretized as $k_j=\dfrac{2\pi}{Na}j, \ j=0,\ldots,N-1$. The Bloch Hamiltonian is found to be of the form (see also \cite{Wang_2008})
\begin{equation}
    \mathcal{H}(k) =
    \begin{bmatrix}
        g(\tilde{k},\phi)+m & 2t_1\cos(\frac{\tilde{k}}{2}) & g(\frac{\tilde{k}}{2},-\phi) & 0 & 0 & 0 &\cdots & 0 \\
        2t_1\cos(\frac{\tilde{k}}{2}) & g(\tilde{k},-\phi)-m & t_1 & g(\frac{\tilde{k}}{2},\phi) & 0 & 0 & \cdots & 0 \\
        g(\frac{\tilde{k}}{2},-\phi) & t_1 & g(\tilde{k},\phi)+m & 2t_1\cos(\frac{\tilde{k}}{2}) & g(\frac{\tilde{k}}{2},-\phi) & 0 &\cdots & 0 \\
        0& g(\frac{\tilde{k}}{2},\phi) & 2t_1\cos(\frac{\tilde{k}}{2}) & g(\tilde{k},-\phi)-m & t_1 & g(\frac{\tilde{k}}{2},\phi) & \ddots & 0 \\
        \vdots & \ddots & \ddots &  \ddots & \ddots & \ddots & \cdots &\vdots
    \end{bmatrix},
    \label{eq:Bloch_matrix}
\end{equation}
with $g(\tilde{k},\phi)=2t_2\cos(\tilde{k}+\phi)$ and $\tilde{k}=ka$.

\begin{figure}[bt]
    \centering
    \includegraphics[width=\linewidth]{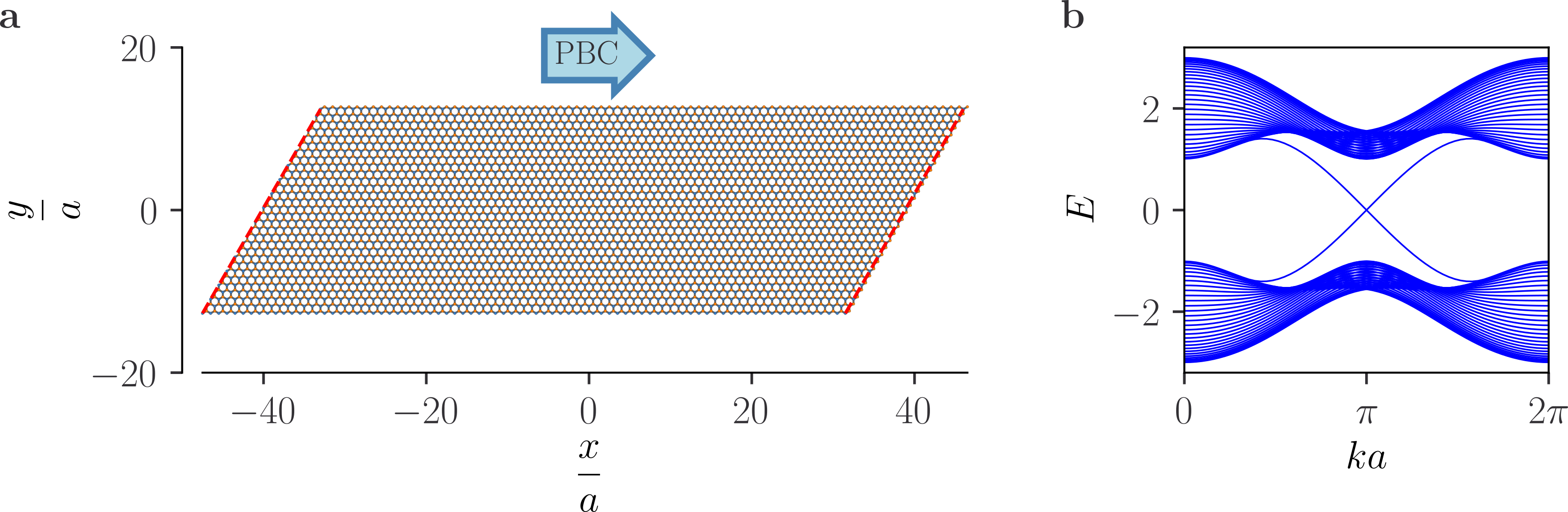}
    \caption{\textbf{Haldane chiral modes in a wide zigzag nanoribbon.}
    \textbf{a} Scheme of the lattice with PBC along the $\mathbf{a}_1$ direction. The dashed red lines at the two ends of the strip mark the sites that are connected by PBC. \textbf{b} Bands of the Haldane model in a strip configuration with zigzag edges, with PBC along the $\mathbf{a}_1$ direction and $N_y=60$ sites in the $y$ (or $\mathbf{a}_2$) direction. The parameters are set as $t_1=1,\ t_2=0.3,\ m=0,\ \phi=\pi/2$.}
    \label{fig:fig1}
\end{figure}

In Fig.~\ref{fig:fig1}\textbf{a} is shown a schematic representation of a strip with zigzag edges, $N_y=60$ sites in the vertical direction, and PBC along the $\mathbf{a}_1$ direction. In Fig.~\ref{fig:fig1}\textbf{b} are reported the corresponding energy bands, obtained via numerical tight-binding diagonalization. Inside the bulk gap are clearly visible the two modes corresponding to the chiral edge states characterizing the topological phase.

It can be expected that by reducing the strip width, the chiral edge states may hybridize because of spatial overlap, giving rise to a gap opening in the edge spectrum. {\color{black} The length scale at which this phenomenon becomes relevant is defined by the decay length of the edge states, that, for the Haldane model on zigzag nanoribbons, has been found to be~\cite{Doh_2013, Cano_2013}
\begin{equation}
    \xi_{\text{loc}}\approx \dfrac{\sqrt{3}}{2}\left[\log\left\{\sqrt{1+\left(\frac{t_1}{4t_2}\right)^2} +\frac{t_1}{4t_2}\right\}\right]^{-1}.
    \label{eq:loc_len}
\end{equation}
This kind of behavior has been explicitly proven} in several contexts~\cite{niu_2008,Ohyama_2011}. In spite of this, an analysis of the topological character of the gapped phases originated by finite size effects is still missing in the context of Chern insulators. In the next section, the results of such a study are presented.

\subsection{Emerging quasi 1D topological phase diagram}
\begin{figure}[bt]
    \centering
    \includegraphics[width=\linewidth]{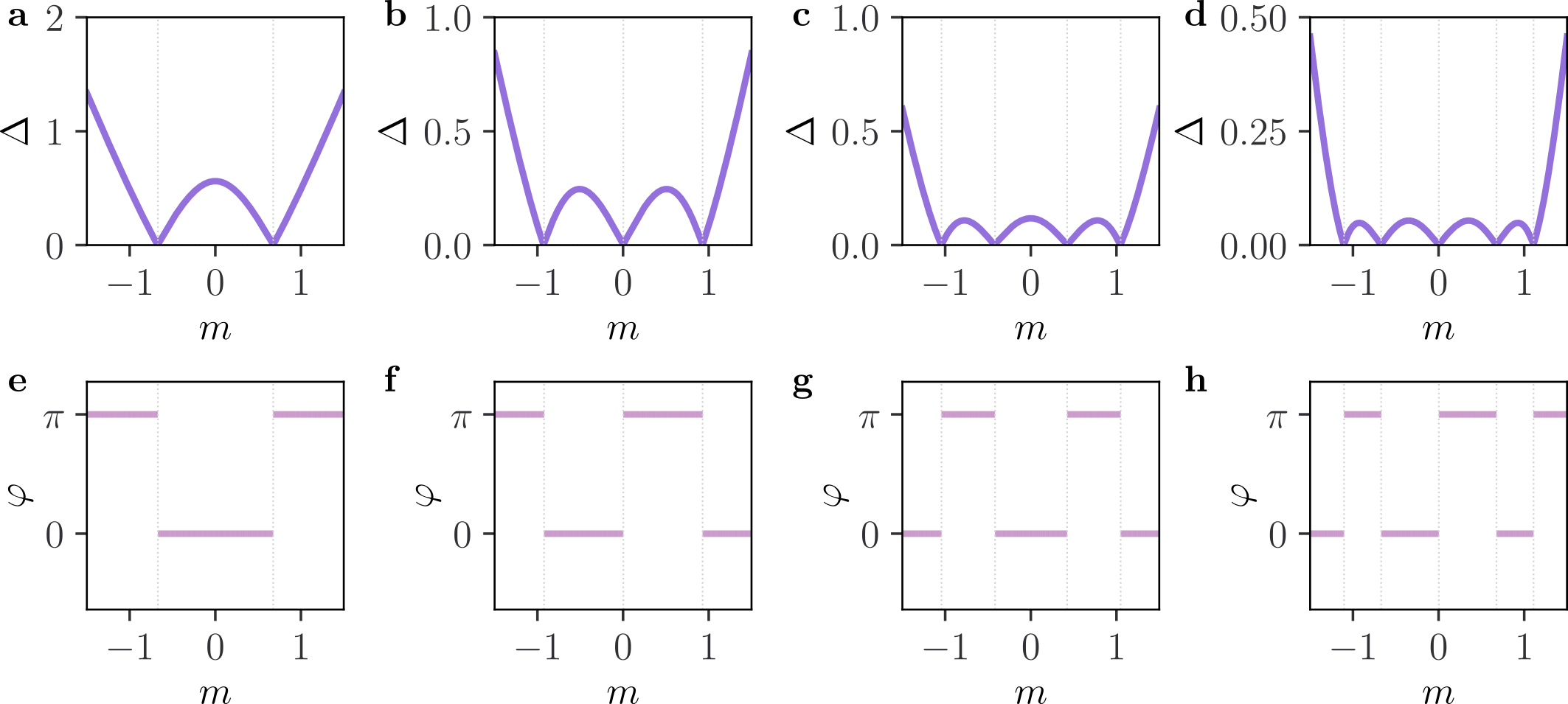}
    \caption{\textbf{Energy gap and Zak phase.}
    Gap width $\Delta$ (top panels) and Zak phase $\varphi$ (bottom panels) as a function of the staggered on-site potential $m$, computed for strips of various widths: (\textbf{a},~\textbf{e}) $N_y=4$;  (\textbf{b},~\textbf{f}) $N_y=6$;  (\textbf{c},~\textbf{g}) $N_y=8$;  (\textbf{d},~\textbf{h}) $N_y=10$. The model parameters are set to $t_1 =1$, $t_2=0.3$ and $\phi=\pi/2$. For $N_y=6$ and $N_y=10$ the system is gapless at $m=0$. In correspondence of every gap closing the Zak phase jumps of $\pi$.}
    \label{fig:Gap-Zak}
\end{figure}

We start by numerically computing the amplitude of the gap $\Delta$ as a function of the staggered on-site potential $m$. We set the energy scale to $t_1 =1$, and we impose $t_2=0.3$ and $\phi=\pi/2$, so that the topological bulk gap of the Haldane model is maximized. We perform our analysis for the Haldane strip with PBC for the different widths $N_y= 4,6,8,10$. The results are reported in Fig.~\ref{fig:Gap-Zak}.

By comparing the top panels (Fig.~\ref{fig:Gap-Zak}\textbf{a}-\textbf{d}) it appears that the number of gap closings and reopenings grows monotonically with the strip width. This non trivial pattern hints to the fact that a topological phase transition may be associated with the gap closings. Interestingly, for the strips whose number of sites in the vertical direction is given by $N_y=4M+2$, the edge spectrum is gapless for $m=0$: in these cases, despite the wave functions of the chiral modes on opposite edges are brought in close proximity, they do not hybridize with each other. This counterintuitive behavior is proven analytically in Supplementary Note 1.

In order to characterize the topology of the zigzag Haldane strips, we use a well established tool for assessing the topological properties of 1D solids: the Zak phase~\cite{Zak_1989}, \textit{i.e.} the natural open-loop extension of the Berry phase~\cite{Berry_1984} when the parameter space is the Brillouin zone. The Zak phase associated with an isolated band was originally defined in terms of the cell-periodic Bloch functions $\ket{u_k}$ as~\cite{Zak_1989}:
\begin{equation}
    \varphi = \mathrm{i} \int_{0}^{2\pi/a}\mathrm{d}k \braket{u_k|\partial_k |u_k}.
\end{equation}
However, the above definition cannot be applied in the present case. Indeed, we are dealing with a multiband system in which the valence bands may cross each other and whose Hamiltonian can only be diagonalized numerically. Thus, we follow the prescription given in~\cite{Resta_1994} for the multi-band case. Given a discretization of the Brillouin zone $k_j=\dfrac{2\pi}{a}\dfrac{j}{N}, \ j \in \{0,\ldots,N-1\}$, for each momentum $k_j$ we compute a basis of eigenstates of the Hamiltonian in Eq.~\ref{eq:Bloch_matrix} for the occupied bands. The resulting $N_y$-dimensional vectors will be denoted as $\ket{u_{nk_0}},\ldots,\ket{u_{nk_{N-1}}}$, $n$ being the band index. Then we enforce the periodic gauge, by defining~\cite{Resta_1994}
\begin{equation}
    \ket{u_{nk_N}}_\ell=\mathrm{e}^{-\mathrm{i}\frac{2\pi}{a}t_\ell}\ket{u_{nk_0}}_\ell,
\end{equation}
where $\ket{u_{nk_j}}_\ell$ is the $\ell$-th component of the eigenvector $\ket{u_{nk_j}}$ and $t_\ell$ the $x$ position of the $\ell$-th site inside the strip unit cell~\cite{yusufaly2013tight}, whose origin we place on the $A$ site at the bottom edge. The Zak phase for the occupied bands is thus defined as~\cite{Resta_1994, vanderbilt_2018, yusufaly2013tight}
\begin{equation}
    \varphi=-\Im \log \det \prod_{j=0}^{N-1}S(k_{j},k_{j+1}),
\end{equation}
with the overlap matrix $S$ given by~\cite{Resta_1994, vanderbilt_2018, yusufaly2013tight}
\begin{equation}
    S(k_{j},k_{j+1})_{mn}=\braket{u_{mk_j}|u_{nk_{j+1}}}.
\end{equation}
This formula is unaffected by any erratic behavior of the phase randomly appended to the eigenvectors by the numerical diagonalization routine~\cite{vanderbilt_2018}. It must be noted that since the Hamiltonian in Eq.~\ref{eq:Bloch_matrix} is real, the Zak phase can only be $0$ or $\pi$. The values obtained for the Zak phase at different values of $m$ are reported in the bottom panels of Fig.~\ref{fig:Gap-Zak}, below the corresponding plots of the energy gap. We find that the Zak phases jump of $\pi$ at each gap closing, confirming that some kind of topological phase transition actually occurs.

In order to gain any insight about which phase is trivial and which is topological we need to make some more physical considerations. As a matter of fact, the Zak phase itself does not have an absolute meaning~\cite{Jens_2017, Atala2013, cayssol_2021}, since its value depends on the choice of the unit cell origin in real space. However, the difference between the values of the Zak phases in two regions of the parameter space is uniquely defined (modulo $2\pi$).

\begin{figure}[bt]
    \centering
    \includegraphics[width=\linewidth]{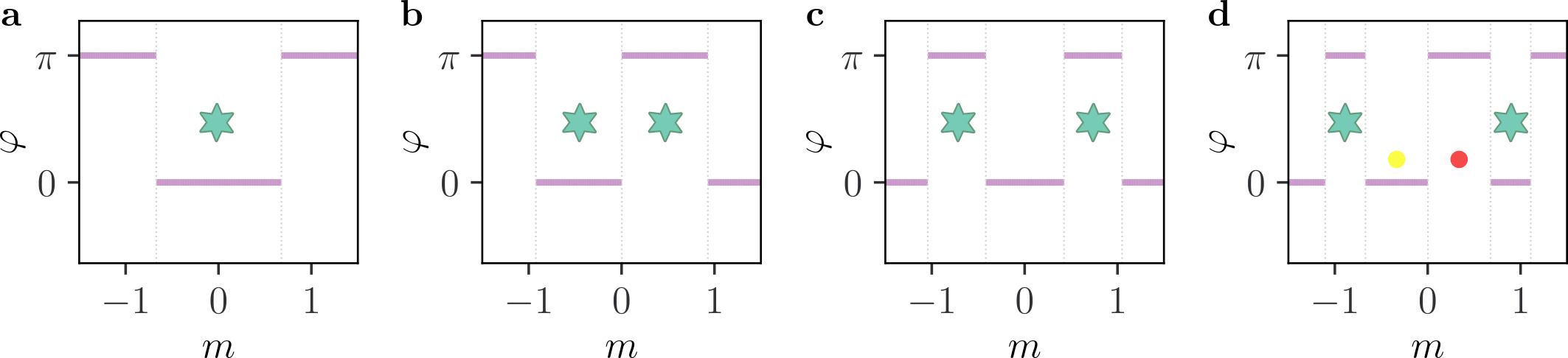}
    \caption{\textbf{Emerging phase diagram.}
    Phase diagram of the model as a function of the staggered on-site potential $m$ for different widths: (\textbf{a}) $N_y=4$;  (\textbf{b}) $N_y=6$;  (\textbf{c}) $N_y=8$; (\textbf{d}) $N_y=10$. The model parameters are set to $t_1 =1$, $t_2=0.3$ and $\phi=\pi/2$. The regions of the parameter space where a star is drawn, are those expected to be topologically non trivial. The yellow and red dots in (\textbf{d}) mark the two phases inspected in Fig.~\ref{fig:jackiw-bs}.}
    \label{fig:phase-diag}
\end{figure}

In the large $m$ limit, the system at half filling is expected to be a trivial insulator: in fact, in this scenario the electrons localize on the sublattice which is lower in energy (depending on the sign of $m$) and the hoppings between sites are suppressed. In view of this, we expect the regions of parameter space characterized by a value of the Zak phase differing by $\pi$ ($0$) from that of the large $m$ limit to be topologically non-trivial (trivial). It is worth pointing out that for $N_y=6$ and $N_y=10$ the limits $m \to \pm \infty$ correspond to different values of the Zak phase. We prove analytically in Supplementary Note 2 that this is a general fact for $N_y=4M+2$. In these cases, the comparison of the Zak phases should be made with the $m\to -\infty$ limit for negative values of $m$ and with the $m\to \infty$ limit for positive ones. The phase diagram emerging from this argument is depicted in Fig.~\ref{fig:phase-diag}, where for each of the widths considered a star is drawn in every region of the parameter space which is expected to be topologically non-trivial. By virtue of bulk boundary correspondence, quasi 0D bound states should occur when the quasi 1D strips are considered under open boundary conditions (OBC) along the $\mathbf{a}_1$ direction and the parameters fall inside one of the topological regions depicted in Fig.~\ref{fig:phase-diag}.

\begin{figure}[bt]
    \centering
    \includegraphics[width=\linewidth]{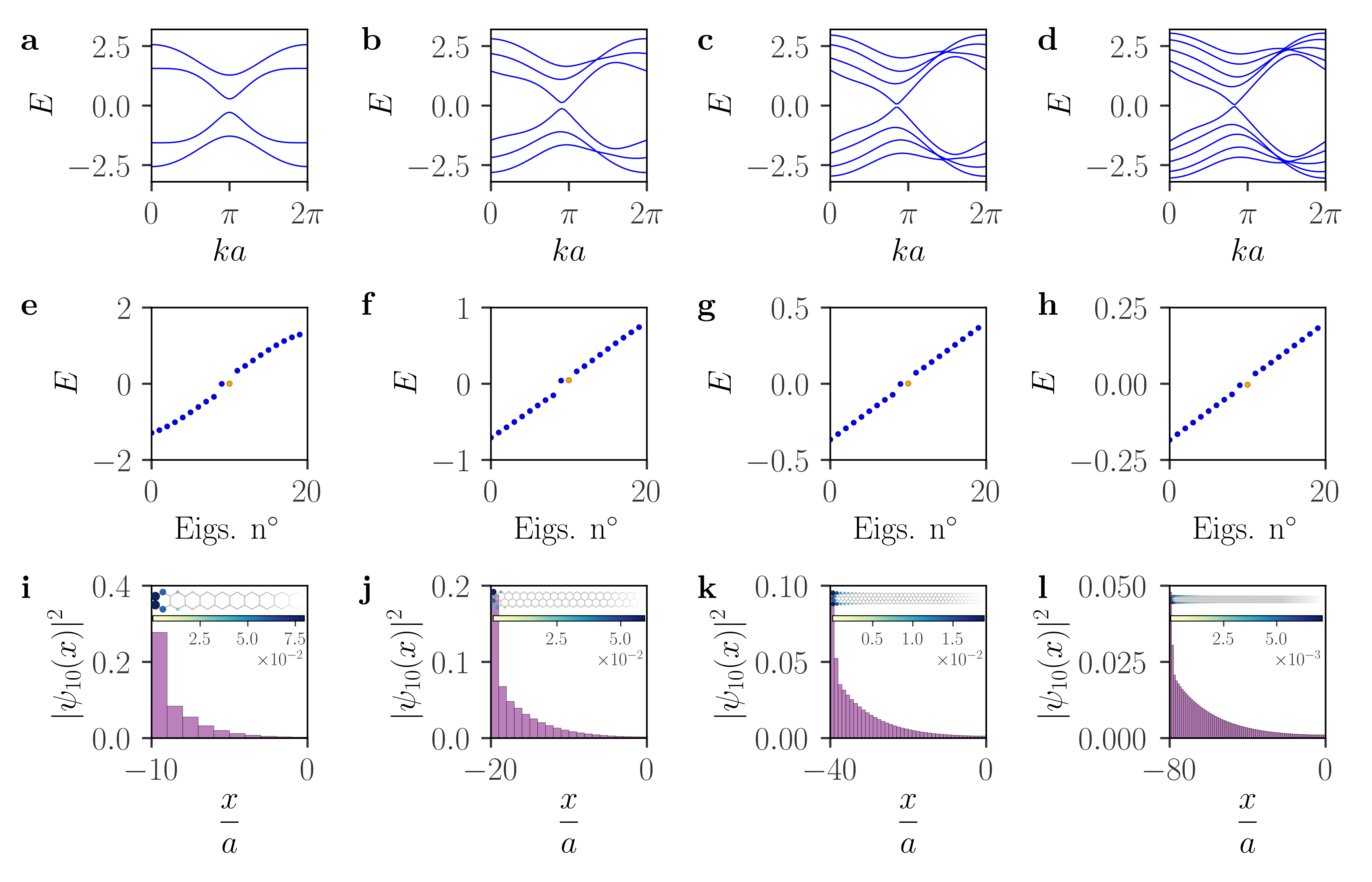}
    \caption{\textbf{Topological quasi 0D bound states.}
    Results from numerical diagonalization in topologically non-trivial regions for different widths: (\textbf{a}, \textbf{e}, \textbf{i}) $N_y=4$;  (\textbf{b}, \textbf{f}, \textbf{j}) $N_y=6$;  (\textbf{c}, \textbf{g}, \textbf{k}) $N_y=8$;  (\textbf{d}, \textbf{h}, \textbf{l}) $N_y=10$. For different values of $m$ for each of the widths considered (respectively $0, \ 0.5, \ 0.8, \ 0.92$), the top row reports the bands of the strips with PBC along $x$ ($\mathbf{a}_1$) and the mid row reports the low energy spectra of the corresponding OBC counterparts, obtained for strips of length $20a$ (\textbf{e}), $40a$ (\textbf{f}), $80a$ (\textbf{g}), $160a$ (\textbf{h}). The values of $m$ chosen for each strip belong to the regions marked with a star in Fig.~\ref{fig:phase-diag}. A doublet of degenerate modes (marked in blue and orange) is visible inside each gap of the OBC spectra: these correspond to bound states localized at the ends of the strips. The 1D probability density profiles of the states associated to the eigenvalue n${}^\circ 10$ (marked in orange) of each of the spectra (\textbf{e}-\textbf{h}) are reported in the bottom row (\textbf{i}-\textbf{l}) as a function of the position for the left half of the various strips ($x<0$). For each of these plots, the insets show the unprojected probability densities. The model parameters are set to $t_1 =1$, $t_2=0.3$ and $\phi=\pi/2$.}
    \label{fig:bound_states}
\end{figure}

To check our predictions we perform numerical diagonalization of the model in an uncompactified geometry along the $\mathbf{a}_1$ direction and we inspect the low energy spectra in the different regions of the parameter space. The OBC strips are cropped with a rectangular geometry, with the long edges parallel to $\mathbf{a}_1$. In Fig.~\ref{fig:bound_states}, for each of the widths taken into account, the bands in PBC geometry (Panels \textbf{a}-\textbf{d}) and the corresponding low energy spectra in the OBC case (Panels \textbf{e}-\textbf{h}) are reported. The values of $m$ at which the diagonalization was performed for each strip, were chosen close to the point of the topological regions where the gap was maximum (cf. Fig.~\ref{fig:Gap-Zak}\textbf{a}-\textbf{d}). The spectra of the finite size systems present two degenerate eigenvalues located inside the gap. In Panels \textbf{i}-\textbf{l} we report the 1D profile of the probability density distributions corresponding to the eigenvalue n${}^\circ 10$ (marked in orange) of the spectra in Panels \textbf{e}-\textbf{h} respectively as a function of the position, together with insets representing the corresponding unprojected probability densities directly on the strips. From the localization pattern of these states, we conclude that the in-gap doublets characterizing the topological phases depicted in Fig.~\ref{fig:phase-diag} correspond to quasi zero-dimensional bound states exponentially localized at the two ends of the strips. We numerically checked the robustness of these 0D bound states against the introduction of random on-site disorder finding that, though their energy is inevitably slightly shifted, they survive as long as the disorder strength does not become comparable with the gap width (see Supplementary Note 3).

The results just discussed prove that the dimensional reduction of the Haldane model, when operated on a strip with zigzag edges, gives rise to a reentrant topological phase diagram, characterized by the emergence of degenerate doublets of in-gap 0D end states. Quite curiously, such phenomenology has no counterpart in the case of armchair nanoribbons, at least in the parameter range we numerically inspected. A qualitative motivation of this peculiar asymmetry is given by the end of the next subsection, where we interpret our results through an effective model.

\subsection{Low energy theory}
In order to understand our findings, we frame them into an effective low energy theory. The minimal low energy model describing the coupling between the two counter-propagating edge states in a zigzag Haldane nanoribbon would be
\begin{equation}
    H_{\text{edge}} = [v_F^{\text{ZZ}}(k-\pi)+\Tilde{m}]\tau_z+\mathcal{M} \tau_x,
    \label{eq:low_en_ham}
\end{equation}
where~\cite{Doh_2013}
\[
    v_{F}^{\text{ZZ}} \approx \dfrac{6 t_2 t_1}{\sqrt{t_1^2+16 t_2^2}}, \qquad \Tilde{m}\approx\dfrac{ m t_1}{\sqrt{t_1^2+16 t_2^2}}.
\]
Unfortunately, here $\mathcal{M}$ is an effective coupling induced by the spatial proximity between the two zigzag edges in the thin strip limit. As such, in principle it depends in a non trivial way on all of the model parameters, as well as on the Bloch momentum $k$.
This severely hinders the derivation of a reliable low energy theory for the present model. Nevertheless, we can still achieve some qualitative predictions-- at least for $m=0$ --by the following hand waving argument. To effectively describe the coupling of the chiral edge states, we consider two 1D chains, representing the two edges of the zigzag nanoribbon. We assume that the sites of the two chains are connected via a coupling which decreases exponentially with the distance, as it is usually done when modelling proximity effects. However, we add a crucial physical input: We introduce a sharp cutoff in the hopping range, effectively coupling only sites that in the actual nanoribbon are connected by the minimum amount of first neighbor hoppings. The motivation behind this choice, is that the coupling between any other pair of edge sites would represent a higher order (negligible) correction.

Omitting the details of the derivation, for which we refer the interested reader to Supplementary Note 4, we find for the two classes considered ($N_y=4M$ or $N_y=4M+2$) and assuming $m=0$
\begin{align}
    &N_y=4M: && \mathcal{M}^{\text{teo}}(k) = \Tilde{\Delta} [1 + \sum_{j=1}^{M} 2\cos(kj)\mathrm{e}^{-(\sqrt{w^2+j^2}-w)/\xi}],
    \label{eq:delta_even}
    \\
    &N_y=4M+2: && \mathcal{M}^{\text{teo}}(k) = \Tilde{\Delta} \sum_{j=0}^{M} 2\cos(k(j+1/2)) \mathrm{e}^{-(\sqrt{w^2+(j+1/2)^2}-w)/\xi},
    \label{eq:delta_odd}
\end{align}
where $w=\frac{\sqrt{3}}{2}\frac{N_y}{2}-\frac{1}{\sqrt{3}}$ is the strip width and $\xi$ is of the order of the chiral edge states localization length (Eq.~\ref{eq:loc_len}).

To benchmark our results, we extract the low energy bands directly from the exact numerical diagonalization. Let us denote by $E_{N_y}(m; k)$ the lowest (positive) energy band for a strip of width $N_y$ (all the other parameters of the model fixed). We can assume that, close to the Dirac point, $E_{N_y}(m; k)$ is well described by the spectrum of the low energy model in Eq.~\ref{eq:low_en_ham}, with a certain unknown function $\mathcal{M}$. Thus, noting that the hybridization between the chiral modes is exponentially suppressed with the strip width-- \textit{i.e.} $\mathcal{M} \xrightarrow[]{N_y\to \infty} 0$ --we can recover (the modulus of) $\mathcal{M}$ for a given $N_y$ as
\begin{equation}
    |\mathcal{M}(m;k)| = \sqrt{E_{N_y}(m; k)^2-E_{\infty}(m; k)^2}.
    \label{eq:mass_numeric}
\end{equation}

\begin{figure}[bt]
    \centering
    \includegraphics[width=\linewidth]{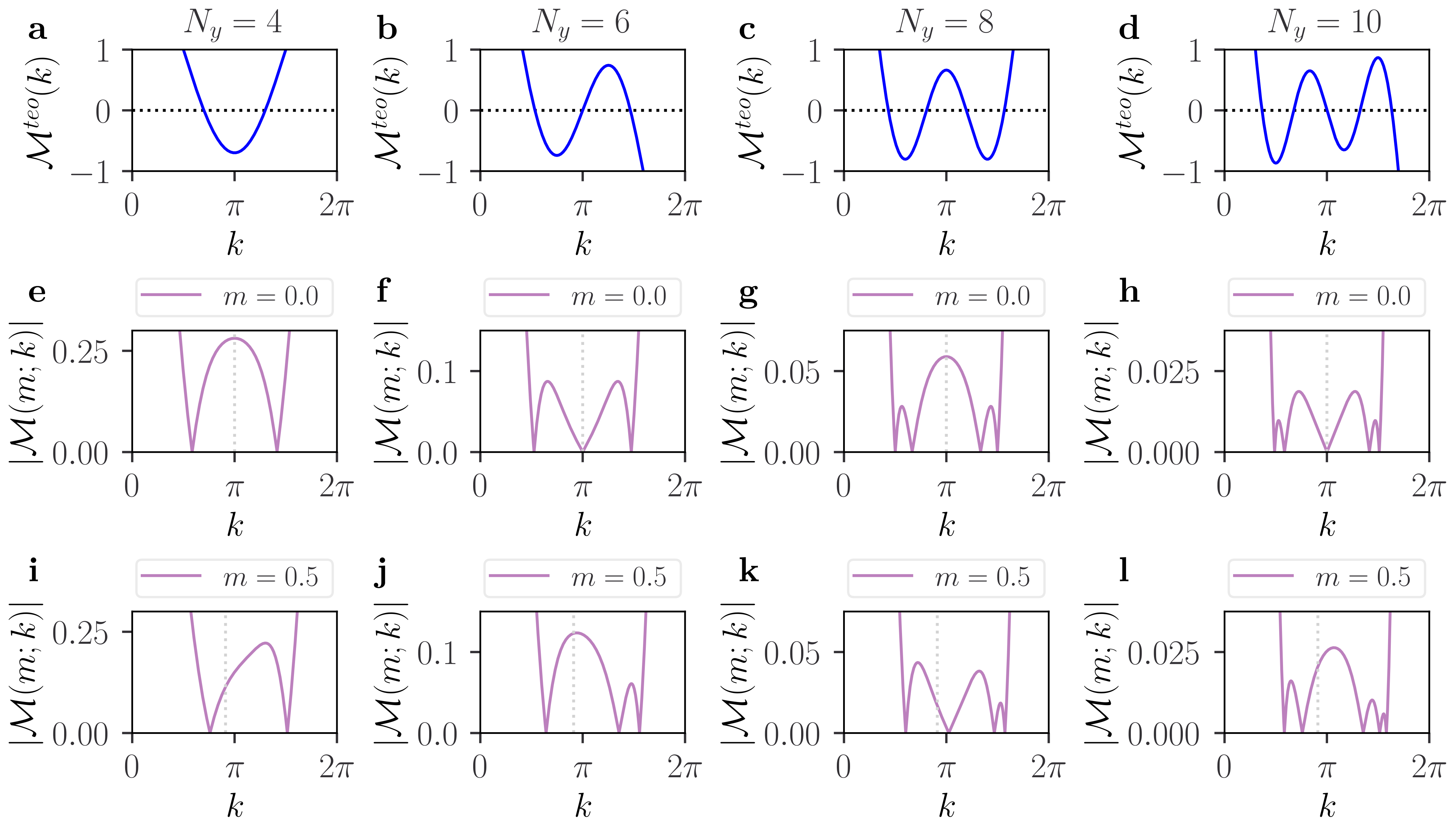}
    \caption{\textbf{Effective toy model validation.}
    Comparison between the effective mass term derived analytically in Eqs.~\ref{eq:delta_even} and \ref{eq:delta_odd}  (\textbf{a}-\textbf{d}) and the ones obtained numerically according to Eq.~\ref{eq:mass_numeric}, for $m=0$ (\textbf{e}-\textbf{h}) and $m=0.5$ (\textbf{i}-\textbf{l}) with the other parameters set to $t_1=1,\ t_2=0.3,\ \phi=\frac{\pi}{2}$. The dashed vertical line in the plots (\textbf{e}-\textbf{l}) indicates the position of the Dirac point, which varies with $m$ according to $k_\text{D} = \pi-\frac{m}{6t_2}$ (cf. Eq.~\ref{eq:low_en_ham}). The $y$-axis in Panels (\textbf{a}-\textbf{d}) are in arbitrary units.}
    \label{fig:mass_teo_vs_num}
\end{figure}

In Fig.~\ref{fig:mass_teo_vs_num}\textbf{a}-\textbf{d} are shown the plots of $\mathcal{M}^{\text{teo}}(k)$ as a function of $k$ for the different widths considered, with $\xi$ set to twice the localization length of the edge states ($\xi_{\text{loc}}$) for $t_2=0.3$ (assuming $t_1=1$). In Panels \textbf{e}-\textbf{h} instead, are reported the corresponding plots of $|\mathcal{M}(0;k)|$, obtained numerically as described in Eq.~\ref{eq:mass_numeric}.
Note that the points in $k$ space where $\mathcal{M}$ stops oscillating and steeply goes up correspond to the points where the edge states of the reference strip ($N_y\to \infty$) merge with the bulk states. Beyond this limit, the low energy Hamiltonian in Eq.~\ref{eq:low_en_ham} is no longer valid, since it does not account for the bulk degrees of freedom.

By comparison between the two rows of plots, one can see that the effective mass term in Eqs.~\ref{eq:delta_even} and \ref{eq:delta_odd} correctly reproduce some qualitative features of the one obtained numerically for $m=0$. More specifically, setting the parameter $t_2$ in the range $[0.1 \sim 0.3]$, we correctly recover the number of nodes of $\mathcal{M}(0;k)$ for the different widths, the fact that the mass term spreads in $k$-space as $t_2$ is increased and, crucially, the fact that $\mathcal{M}(0;\pi)=0$ for $N_y=4M+2$. Further support to these statements can be found in Supplementary Figs.~4, 5 and 6, where, for each of the widths considered, we report the plots of $|\mathcal{M}|$ as a function of $k$ for different values of $m$ and $t_2$. It is worth mentioning that the shape of $\mathcal{M}(0;k)$ more closely resembles that of a sinusoidal-like function for smaller values of $t_2$ ($t_2\sim 0.1-0.2$). The deviations from the sinusoidal pattern predicted by the effective low energy theory observed in Fig.~\ref{fig:mass_teo_vs_num}\textbf{e}-\textbf{l} ($t_2=0.3$), are probably due to the fact that the first has been derived taking into account first neighbor hoppings only. That being the case, it is expected to be more reliable for smaller values of $t_2$.

Furthermore, the numerical results for $\mathcal{M}$ show that, at least at a qualitative level, for $m>0$ the mass term is shifted to the \emph{right} in the Brillouin zone (see Fig.~\ref{fig:mass_teo_vs_num}\textbf{i}-\textbf{l} and Supplementary Figs.~4, 5 and 6). On the other hand, we know that for $m>0$ the Dirac point moves to the \emph{left} (dashed vertical lines in Fig.~\ref{fig:mass_teo_vs_num}\textbf{i}-\textbf{l}). Therefore, there must be a (set of) value(s) of $m$ for which the nodes of $\mathcal{M}(m;k)$ coincide with the shifted Dirac point. This mechanism qualitatively explains the mass inversion in our model and, consequently, the reentrant topological phase diagram retrieved numerically.

Finally, we briefly address the reason why armchair nanoribbons do not share this kind of phenomenology. In the Haldane model on armchair nanoribbons the Dirac point is at $k=0$, and is insensitive to variation of the staggered mass. This can be understood by noticing that, in contrast with the zigzag case, each armchair edge has the same number of $A$ and $B$ sites. Thus, varying $m$ does not shift the energy of the edge states. This hints to the fact that in the case of armchair geometry, the properties of the edge states are less likely to be tuned by varying the staggered mass $m$. Moreover, the edge states in zigzag Haldane nanoribbons present a peculiar property: as shown by Eq.~\ref{eq:loc_len}, $\xi_{\text{loc}}$ grows with $t_2$ in the zigzag case. On the other hand, the Haldane edge states on armchair nanoribbons have their localization length inversely proportional to $t_2$~\cite{Cano_2013}. Thus, in the zigzag case, bigger values of $t_2$ lead both to a wider topological bulk gap in the 2D limit, and to a larger localization length of the edge states. This feature, which is not shared by armchair nanoribbons, elects thin zigzag nanoribbons as the optimal playground to inspect the finite size effects on the Haldane edge states.

\subsection{Jackiw-Rebbi like bound states}
Interestingly, if we consider a setup of our model in which the staggered mass term $m$ interpolates between two values in adjacent regions of the phase diagram with different Zak phases, we observe the occurrence of a bound state, localized at the transition point and with energy lying inside the gap. Thus, leaning on the results from Jackiw and Rebbi~\cite{Jackiw_1976}, we can conclude that such bound state retains a fractional charge of $\pm \frac{e}{2}$~\cite{Qi_2008, Ziani_2020}. Strikingly, the bound state is present even when the mass $m$ interpolates between two regions that, despite having different Zak phases, do not host bound states against the vacuum (more details in Supplementary Note 5).

In Fig.~\ref{fig:jackiw-bs} we report an explicit example for a strip with $N_y=10$ and $L=200a$, in which the on-site staggered potential is set to $\mp 0.3$ for the $A$ and $B$ sites respectively on the left part and to $\pm 0.3$ on the right part (with $m=0$ at the interface to smoothen out the jump). These two regions of the phase diagram are marked with a yellow and red dot respectively in Fig.~\ref{fig:phase-diag}\textbf{d}. All the other parameters of the model are left unchanged. In Panel \textbf{a} is reported a density plot of the on-site potential close to the transition point, where the sign of $m$ switches for the two sublattices. By performing real space numerical diagonalization we obtain the low energy spectrum and the corresponding eigenvectors. The first is shown in Panel \textbf{b}, where an isolated eigenvalue is clearly visible inside the gap: this corresponds to a bound state localized at the point where $m$ switches its sign, as demonstrated by the plot of its probability density in Panel \textbf{c}. From Fig.~\ref{fig:Gap-Zak}\textbf{d}, we see that from $m=-0.3$ to $m=+0.3$ the Zak phase jumps of $\pi$, so that the occurrence of a bound state at the domain wall is actually expected according to our analysis.

\begin{figure}[bt]
    \centering
    \includegraphics[width=\linewidth]{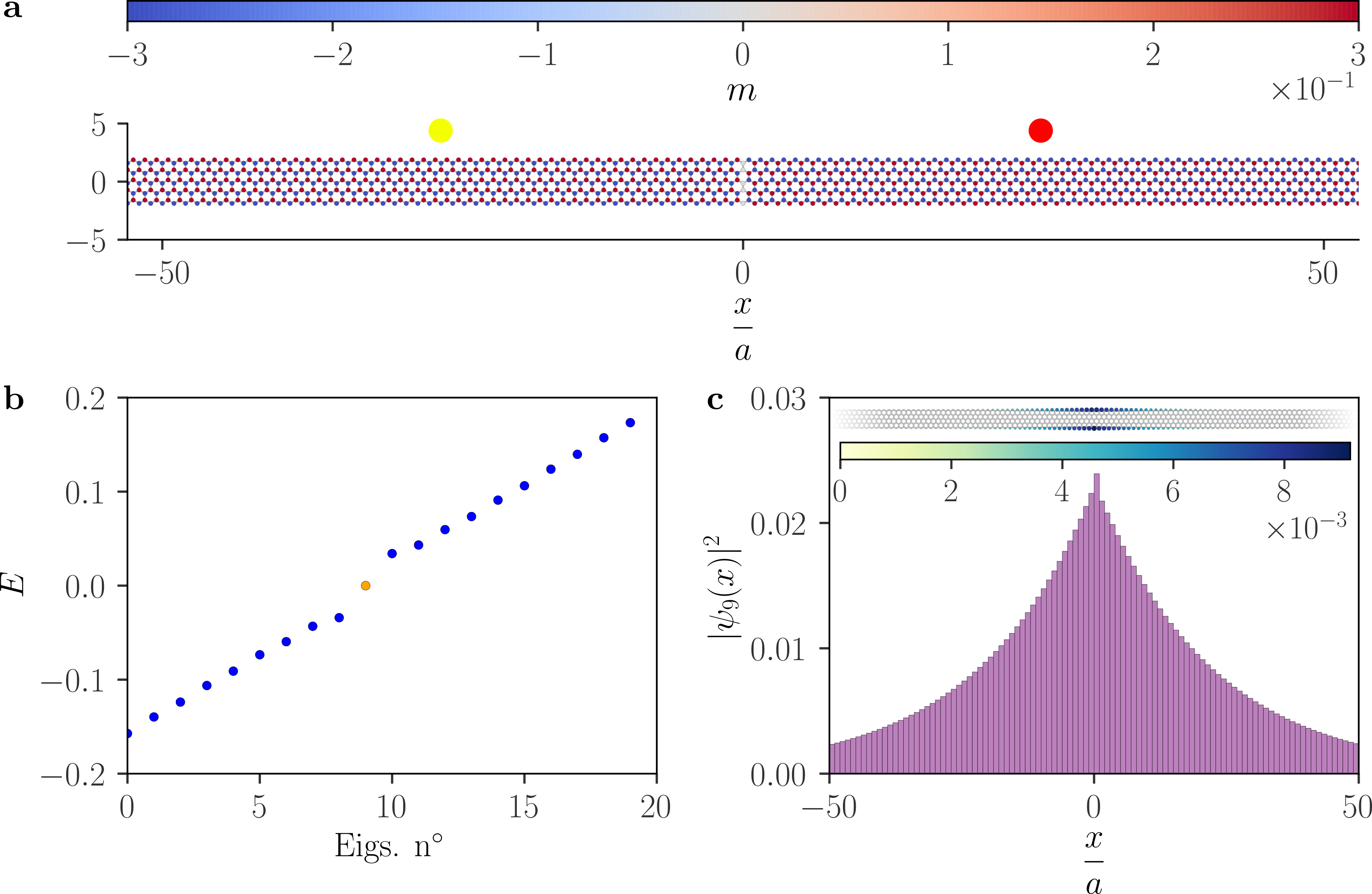}
    \caption{\textbf{Jackiw-Rebbi bound states at domain walls.}
    Results obtained via numerical diagonalization for a strip with $N_y=10$ and $L=200a$ with spatially varying on-site potential. According to Jackiw-Rebbi theory, a bound state is expected in correspondence of mass inversion (which we identify as a jump of $\pi$ in the Zak phase). The model parameters are set to $t_1 =1$, $t_2=0.3$ and $\phi=\pi/2$. \textbf{a} Representation of the on-site potential for the central region of the strip. The left side of the strip ($x<0$) has $m=-0.3$, while the right side ($x>0$) has $m=0.3$ (cf. yellow and red dot of Fig.~\ref{fig:phase-diag}\textbf{d}). The Zak phases in the two regions differ of $\pi$.
    \textbf{b} Low energy spectrum obtained via numerical diagonalization. An isolated eigenvalue is visible inside the gap (marked in orange). \textbf{c} 1D profile of the probability density of the state associated with the in-gap eigenvalue for $-50a<x<50a$, with an inset showing the corresponding unprojected probability distribution directly on the strip. The state appears to be exponentially localized at the domain wall.}
    \label{fig:jackiw-bs}
\end{figure}

\section{Discussion}\label{sec:Disc}

In this paper we have studied the role of finite size effects on the Haldane model. We have shown that, in the case of zigzag strip geometry, the chiral edge modes can, as expected, hybridize and develop a gap. Surprisingly, however, such gap can be both trivial or topological in the sense that, in the uncompactified geometry, bound states can be present or absent depending on the value of the trivial mass, even in the topological phase of the Haldane model. In other words, we unveiled a phase diagram which presents a width-dependent reentrant behavior with respect to the tuning of the on-site staggered potential. Moreover, we have reinterpreted such reentrant structure within an effective minimal model describing the coupling between the chiral edge states of the Haldane model in zigzag nanoribbons.

We have then proven that, when present, the bound states are robust against on-site random disorder. Besides, we have shown that they also occur in correspondence of domain walls in the on-site staggered potential and, consequently, that they bear a fractional charge of $\pm\frac{e}{2}$. 
The topological nature of the bound states is witnessed by the behavior of the Zak phase. Indeed, we can hence conclude that the mass associated to the tunneling between the edges competes in a definitely non-trivial fashion with the other masses of the model, generating a rich phenomenology.

The implications of our results are diverse. In the context of two-dimensional topological insulators and Chern insulators they imply, for instance, that the transport properties of setups where constrictions are present might be affected by the presence of zero modes, and hence show resonances. Moreover, given that a similar gap structure also characterizes the 2D Kitaev model~\cite{Potter_2010,Shen_2011} for topological superconductivity, the impact of our results bears also consequences on the field of Majorana zero modes~\cite{Kitaev2001, Albrecht_2016, Mourik_2012} and parafermions~\cite{Fendley_2012, Alicea_2016, calzona_2018}, paving the way to new possibilities for implementing such non-abelian excitations.

\section{Methods}\label{sec:Methods}
\subsection{Numerical diagonalization tools}
The finite size model construction and the numerical diagonalization have been performed using the package Pybinding~\cite{dean_moldovan_2020_4010216}.
\subsection{Numerical computation of the Zak phase}
The computation of the gaps and of the Zak phase as a function of the staggered on-site potential have been performed with an original code and the results for the Zak phase have been benchmarked with existing packages.
\backmatter

\section*{Declarations}

\bmhead{Data availability}
All data relevant to the paper are reported in the main text and in the Supplementary Information. All the numerically generated points reported in the plots of this paper and in the accompanying supplementary are obtained as described in section \ref{sec:Methods}. The actual codes used to produce the results reported in this paper and in the accompanying supplementary are available from the corresponding author upon request.

\bmhead{Acknowledgments}
N.T.Z. acknowledges the funding through the NextGenerationEu Curiosity Driven Project ``Understanding even-odd criticality''. N.T.Z. and M.S. acknowledge the funding through the ``Non-reciprocal supercurrent and topological transitions in hybrid Nb-InSb nanoflags'' project (Prot. 2022PH852L) in the framework of PRIN 2022 initiative of the Italian Ministry of University (MUR) for the National Research Program (PNR).

\bmhead{Author contributions}
Conceptualization, M.S. and N.T.Z. Development and preparation of the first draft S.T. All authors contributed to the interpretation of the results and reviewed the manuscript.

\bmhead{Competing interests}
The authors declare no competing interests.

\bmhead{Supplementary information}
Details of the calculations and further numerical results can be found in the accompanying Supplementary Information.


\bibliography{biblio}

\end{document}


\title[Supplementary]{Supplementary Information}


\author*[1]{\fnm{Simone} \sur{Traverso}}\email{simone.traverso@edu.unige.it}

\author[1,2]{\fnm{Maura} \sur{Sassetti}}\email{sassetti@fisica.unige.it}

\author[1,2]{\fnm{Niccol\`o Traverso} \sur{Ziani}}\email{traversoziani@fisica.unige.it}

\affil*[1]{\orgdiv{Physics Department}, \orgname{University of Genoa}, \orgaddress{\street{Via Dodecaneso 33}, \city{Genova}, \postcode{16146}, \country{Italia}}}

\affil[2]{\orgname{CNR-SPIN}, \orgaddress{\street{Via Dodecaneso 33}, \city{Genova}, \postcode{16146}, \country{Italia}}}

\keywords{topological insulators, Haldane model, bound states, Zak phase}



\maketitle

\section*{Supplementary Figures}
\begin{figure}[p]
    \centering
    \includegraphics[width=\linewidth]{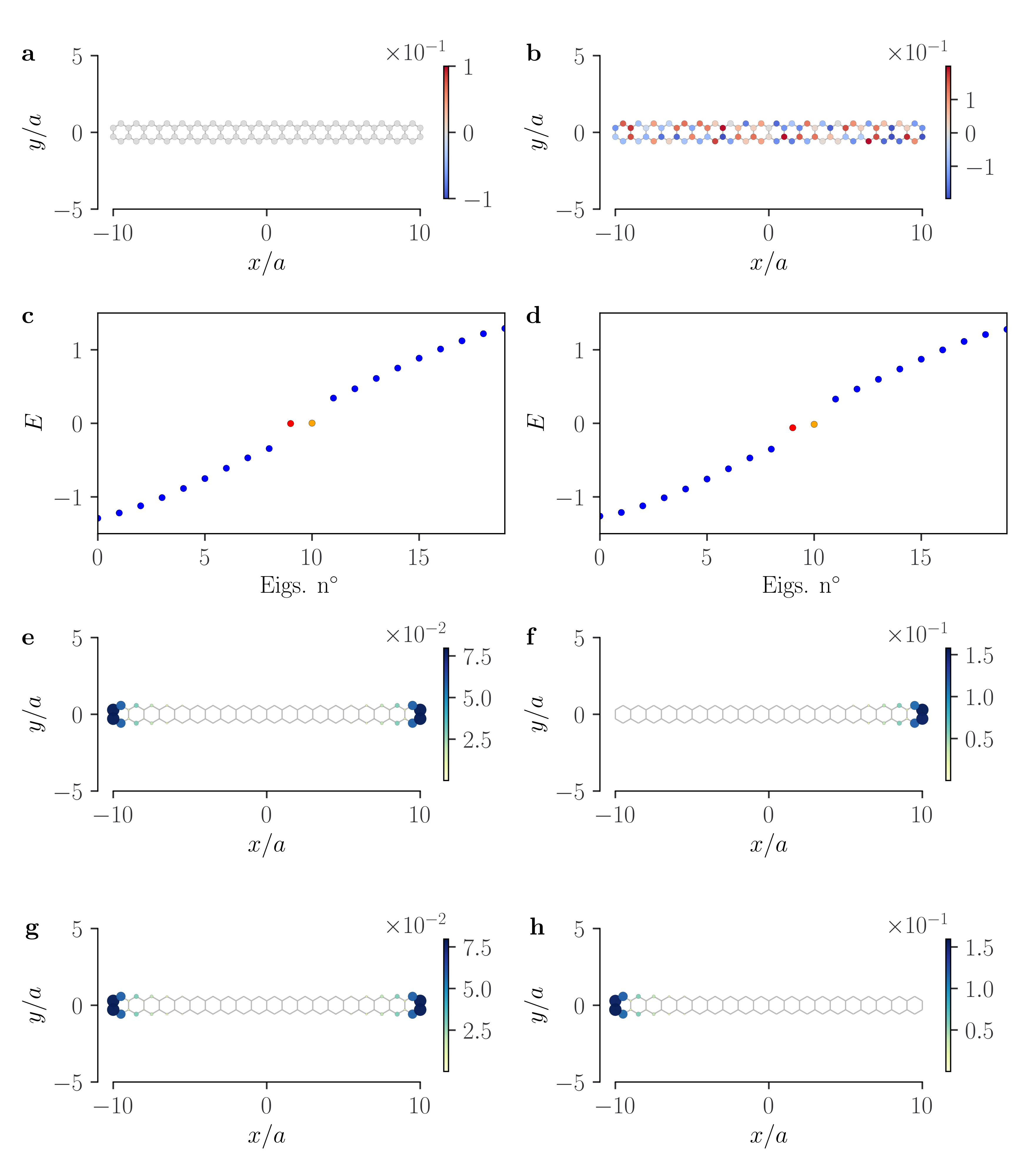}
    \caption{Comparison between the results obtained through numerical diagonalization for the Haldane model on a zigzag nanoribbon (OBC) with $N_y=4$, $L=20a$ and $m=0$, with (Panels \textbf{b}, \textbf{d}, \textbf{f}, \textbf{h}) and without (Panels \textbf{a}, \textbf{c}, \textbf{e}, \textbf{g}) the addition of a random on-site potential. The model parameters are set to $t_1=1, \ t_2=0.3, \ \phi=\frac{\pi}{2}, \ V=0.2$. Panels \textbf{a}, \textbf{b}: representation of the on-site potential on the strip.  Panels \textbf{c}, \textbf{d}: low energy spectra respectively without and with random noise. Though it may be difficult to discern from the plot, the in-gap eigenvalues in Panel \textbf{d} (marked in red and orange) are slightly split because of the noise. Panels \textbf{e}, \textbf{g}: probability density plots of the eigenstates associated to the eigenvalues highlighted respectively in red and orange in Panel \textbf{c}; in absence of noise, the eigenstates show equal localization on both ends of the strip. Panels \textbf{f}, \textbf{h}: probability density plots of the eigenstates associated to the eigenvalues highlighted respectively in red and orange in Panel \textbf{d}; each of the two eigenstates is localized just on one end, as a consequence of the noise induced splitting between the bound state energies.}
    \label{fig:noise_N_2}
\end{figure}

\begin{figure}[p]
    \centering
    \includegraphics[width=\linewidth]{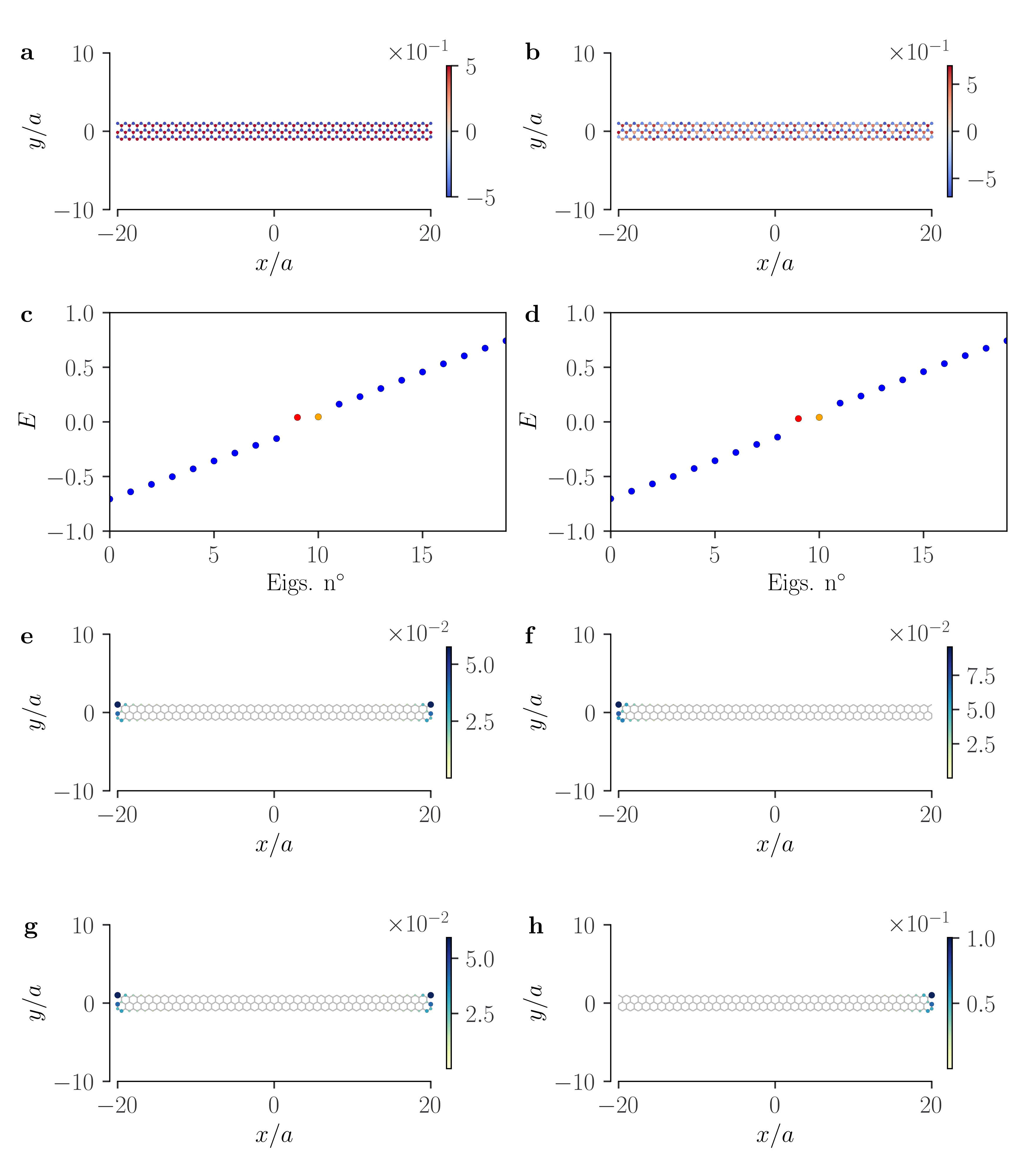}
    \caption{Comparison between the results obtained through numerical diagonalization for the Haldane model on a zigzag nanoribbon (OBC) with $N_y=6$, $L=40a$ and $m=0.5$, with (Panels \textbf{b}, \textbf{d}, \textbf{f}, \textbf{h}) and without (Panels \textbf{a}, \textbf{c}, \textbf{e}, \textbf{g}) the addition of a random on-site potential. The model parameters are set to $t_1=1, \ t_2=0.3, \ \phi=\frac{\pi}{2}, \ V=0.2$. Panels \textbf{a}, \textbf{b}: representation of the on-site potential on the strip.  Panels \textbf{c}, \textbf{d}: low energy spectra respectively without and with random noise. Though it may be difficult to discern from the plot, the in-gap eigenvalues in Panel \textbf{d} (marked in red and orange) are slightly split because of the noise. Panels \textbf{e}, \textbf{g}: probability density plots of the eigenstates associated to the eigenvalues highlighted respectively in red and orange in Panel \textbf{c}; in absence of noise, the eigenstates show equal localization on both ends of the strip. Panels \textbf{f}, \textbf{h}: probability density plots of the eigenstates associated to the eigenvalues highlighted respectively in red and orange in Panel \textbf{d}; each of the two eigenstates is localized just on one end, as a consequence of the noise induced splitting between the bound state energies.}
    \label{fig:noise_N_3}
\end{figure}

\begin{figure}[p]
    \centering
    \includegraphics[width=\linewidth]{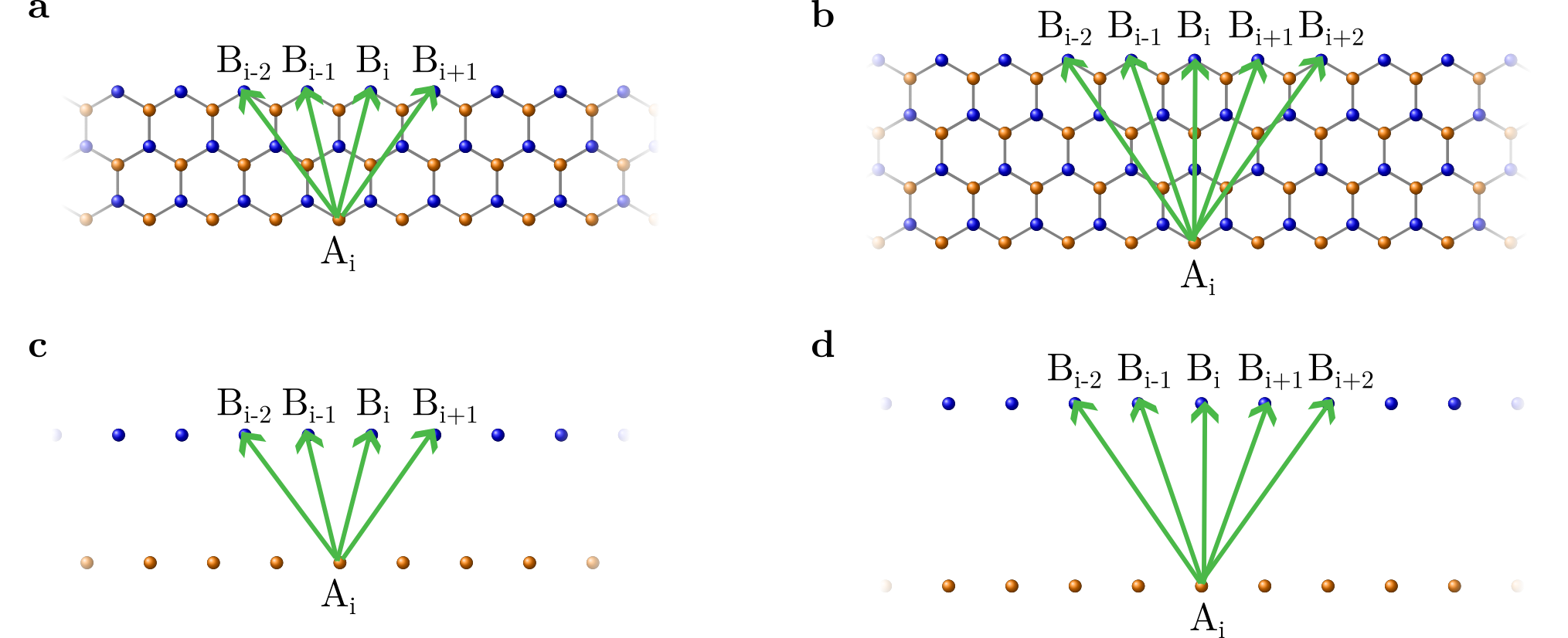}
    \caption{\color{black} First-neighbor coupling between the sites of the two edges for a strip of the family $N_y=4M+2$ and of the family $N_y=4M$ in Panel \textbf{a} and \textbf{b} respectively. In Panel \textbf{c} and \textbf{d} the corresponding effective 1D chains.}
    \label{fig:eff_coupl_scheme}
\end{figure}

\begin{figure}[p]
    \centering
    \includegraphics[width=\linewidth]{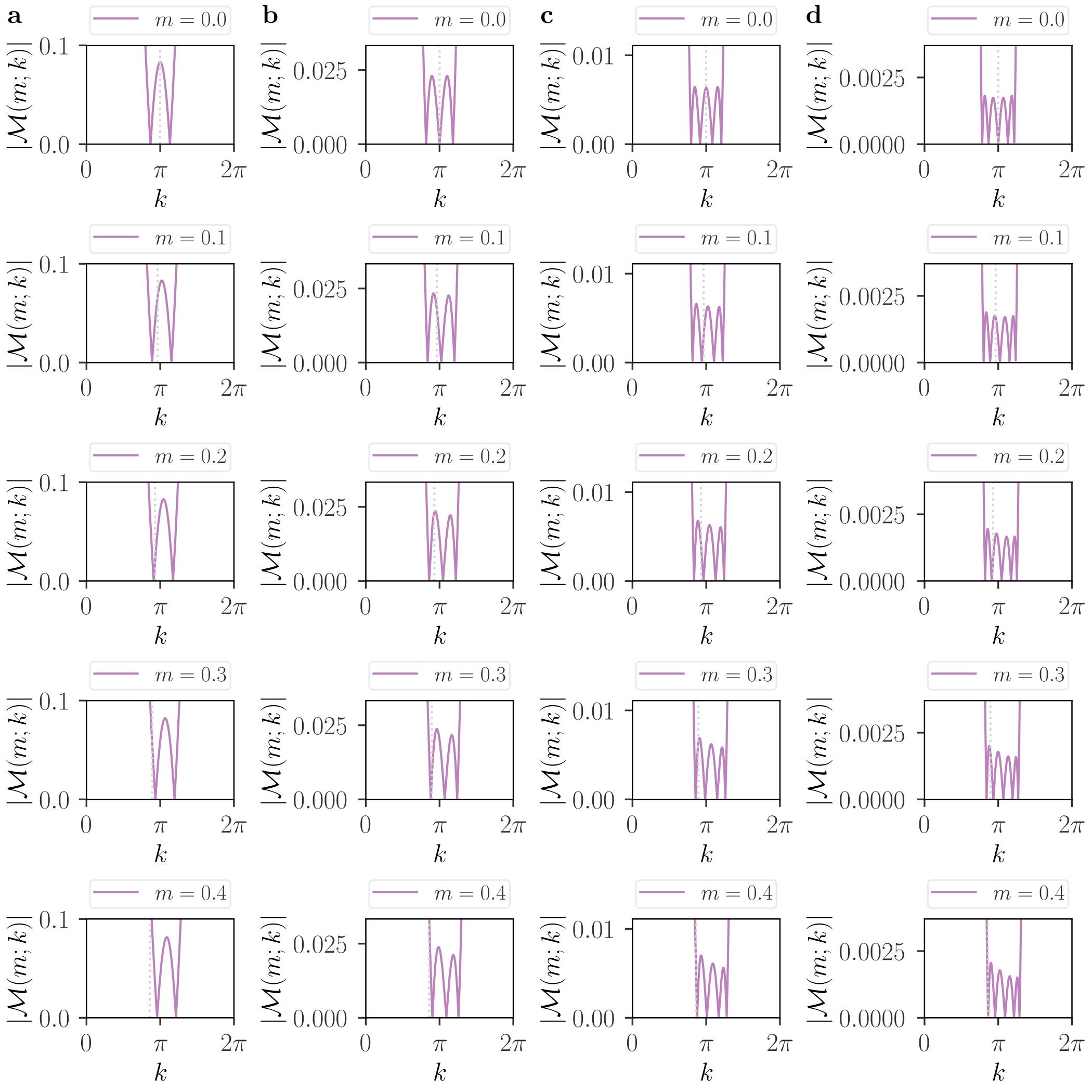}
    \caption{\color{black} Plots of the effective mass term (modulus), obtained numerically as described in Eq.~\ref{eq:mass_num}. The model parameters are set as $t_1=1, \phi = \pi/2$ and $t_2=0.15$. In column \textbf{a} is shown $|\mathcal{M}(m;k)|$ for a strip with $N_y=4$ for increasing values of $m$ (indicated in the lable). The dashed vertical line indicates the position of the Dirac point as a function of the staggered mass. In columns \textbf{b}, \textbf{c}, \textbf{d} are reported the corresponding plots for strips with $N_y=6$, $N_y=8$ and $N_y=10$ respectively.}
    \label{fig:Mk_t2_015-num}
\end{figure}

\begin{figure}[p]
    \centering
    \includegraphics[width=\linewidth]{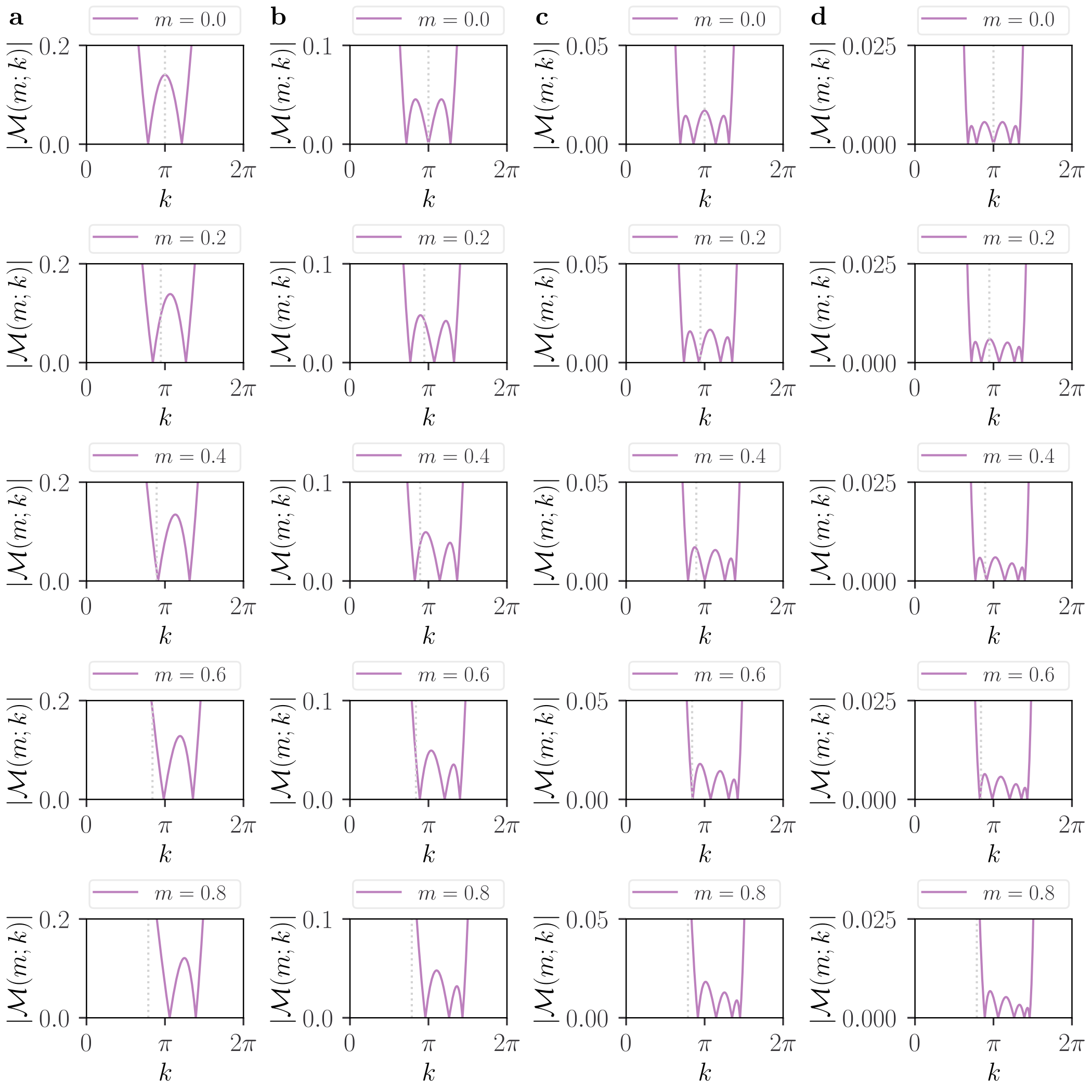}
    \caption{\color{black} Plots of the effective mass term (modulus), obtained numerically as described in Eq.~\ref{eq:mass_num}. The model parameters are set as $t_1=1, \phi = \pi/2$ and $t_2=0.2$. In column \textbf{a} is shown $|\mathcal{M}(m;k)|$ for a strip with $N_y=4$ for increasing values of $m$ (indicated in the lable). The dashed vertical line indicates the position of the Dirac point as a function of the staggered mass. In columns \textbf{b}, \textbf{c}, \textbf{d} are reported the corresponding plots for strips with $N_y=6$, $N_y=8$ and $N_y=10$ respectively.}
    \label{fig:Mk_t2_02-num}
\end{figure}

\begin{figure}[p]
    \centering
    \includegraphics[width=\linewidth]{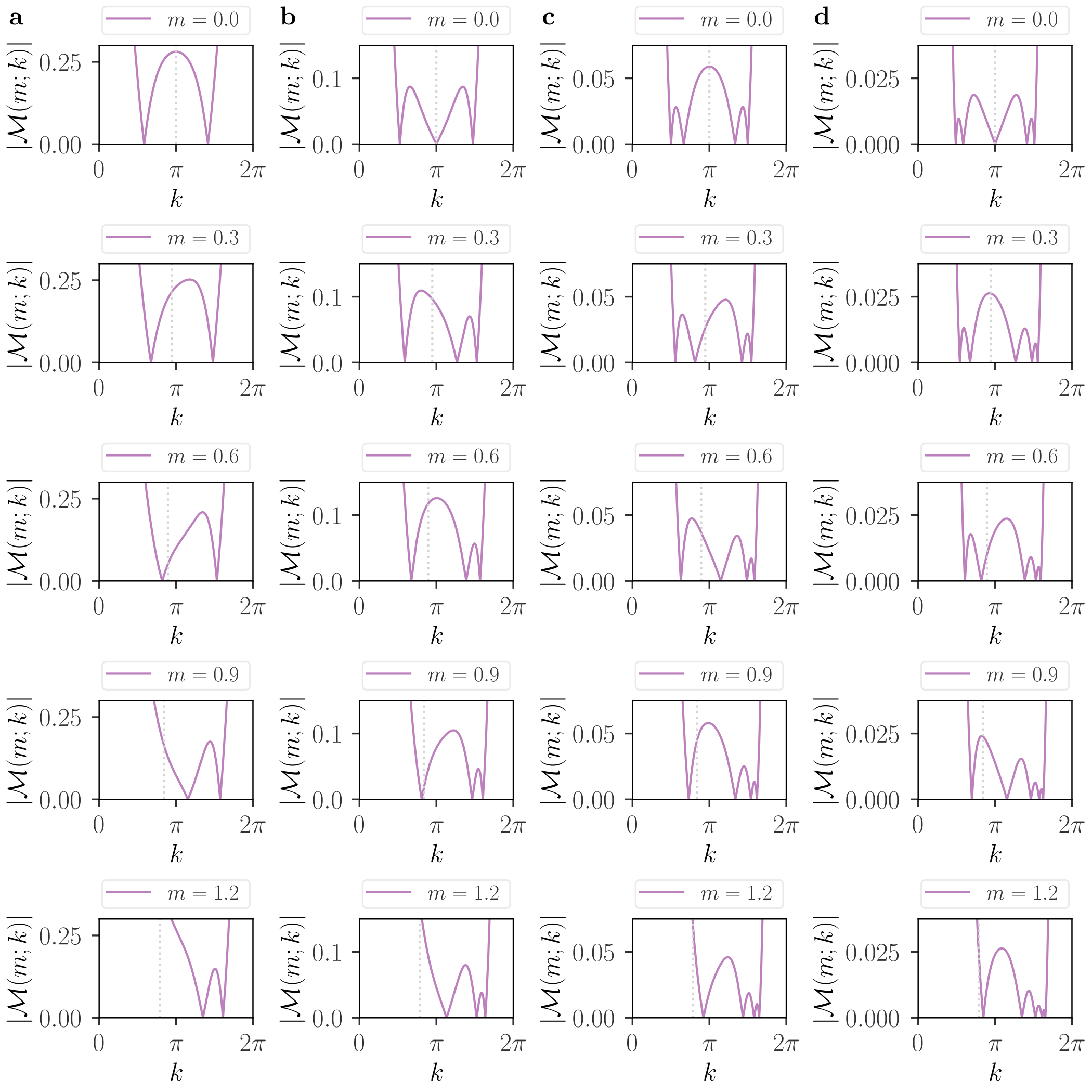}
    \caption{\color{black} Plots of the effective mass term (modulus), obtained numerically as described in Eq.~\ref{eq:mass_num}. The model parameters are set as $t_1=1, \phi = \pi/2$ and $t_2=0.3$. In column \textbf{a} is shown $|\mathcal{M}(m;k)|$ for a strip with $N_y=4$ for increasing values of $m$ (indicated in the lable). The dashed vertical line indicates the position of the Dirac point as a function of the staggered mass. In columns \textbf{b}, \textbf{c}, \textbf{d} are reported the corresponding plots for strips with $N_y=6$, $N_y=8$ and $N_y=10$ respectively.}
    \label{fig:Mk_t2_03-num}
\end{figure}

\begin{figure}[p]
    \centering
    \includegraphics[width=\linewidth]{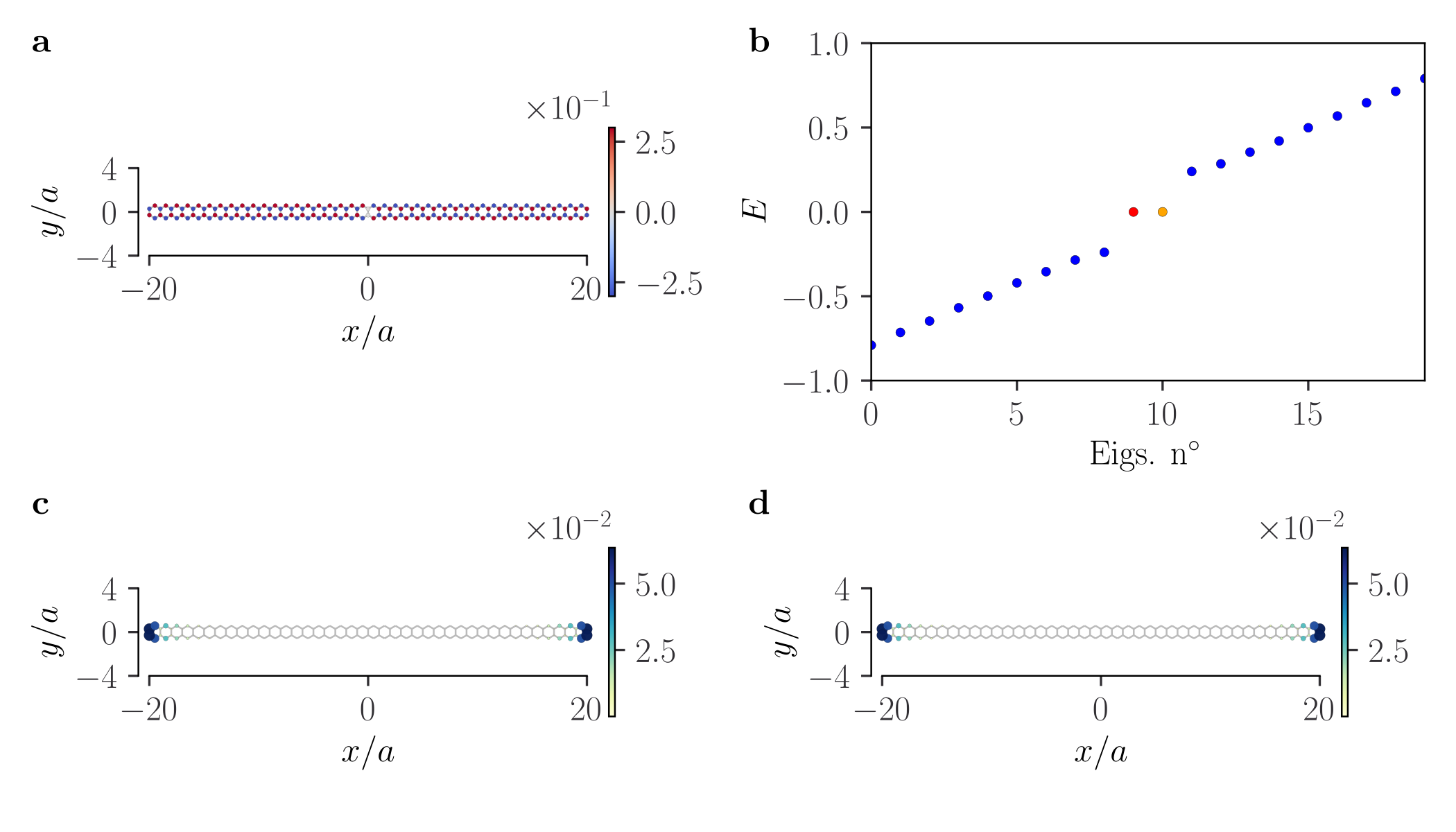}
    \caption{Results obtained through numerical diagonalization for the Haldane model on a zigzag nanoribbon (OBC) with $N_y=4$ and $L=40a$ with the staggered on-site potential interpolating between $-0.3$ and $0.3$. Panel \textbf{a}: representation of the on-site potential, switching its sign from $x<0$ to $x>0$. According to the phase diagram in the main text the two values of $m$ pertain to the same region, which is topologically non trivial in terms of end states. Panel \textbf{b}: low energy spectrum of the model. The gap hosts two degenerate eigenvalues. Panel \textbf{c},\textbf{d}: probability density plots of the states corresponding to the eigenvalues n${}^\circ 9$ and n${}^\circ 10$ in Panel \textbf{c} respectively (cf. red and orange dot). The states are equally localized on the two ends of the strip.}
    \label{fig:VarPot_N_4-M_-0.3_0.3}
\end{figure}

\begin{figure}[p]
    \centering
    \includegraphics[width=\linewidth]{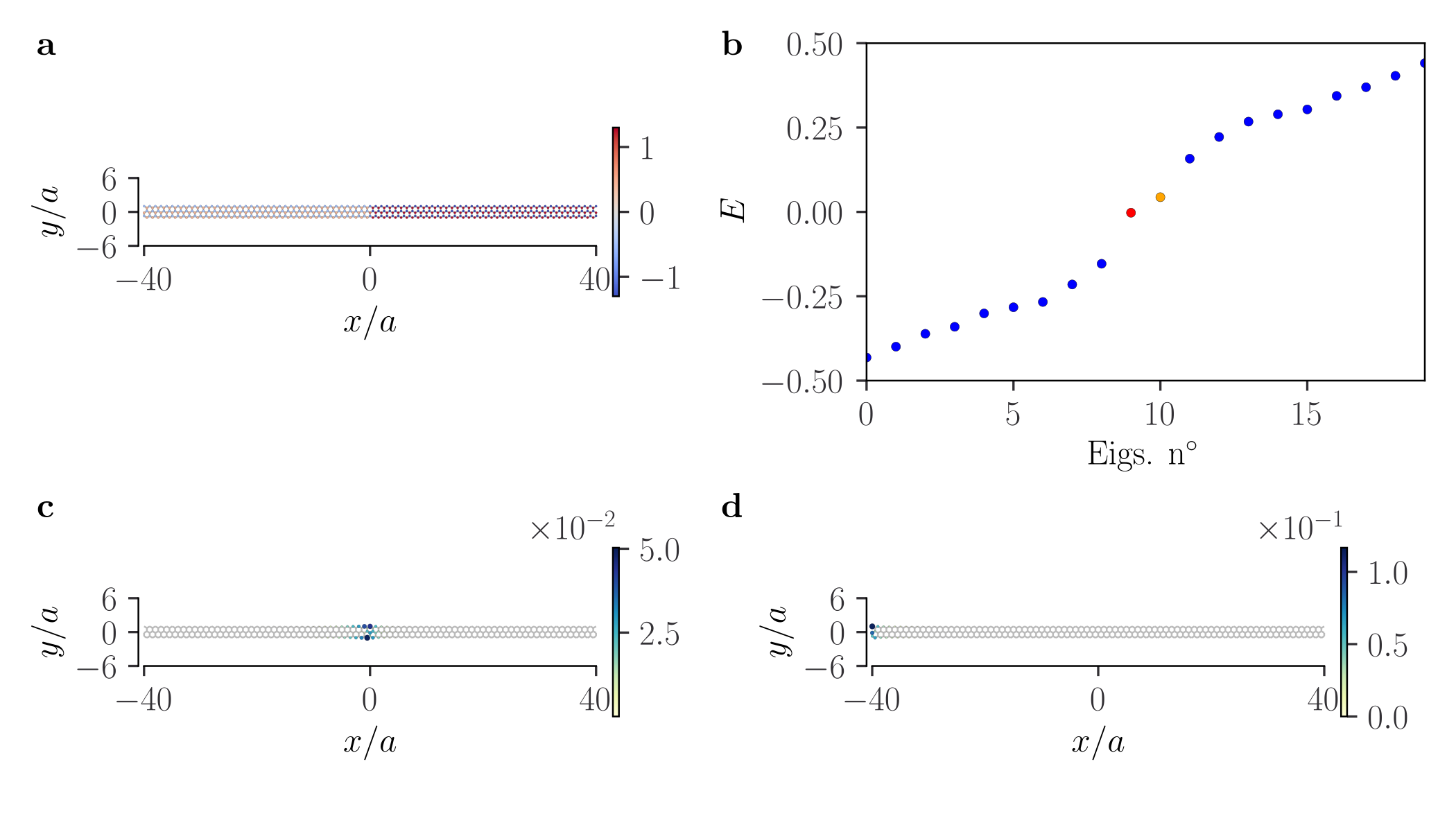}
    \caption{Results obtained through numerical diagonalization for the Haldane model on a zigzag nanoribbon (OBC) with $N_y=6$ and $L=80a$ with the staggered on-site potential interpolating between $0.5$ and $1.3$. Panel \textbf{a}: representation of the on-site potential, changing from $m=0.5$ for $x<0$ to $m=1.3$ for $x>0$. According to the phase diagram in the main text the two values of $m$ pertain to the two distinct contiguous regions, having the Zak phase differing of $\pi$. Only the one containing $m=0.5$ is topologically non trivial in terms of end states. Panel \textbf{b}: low energy spectrum of the model. The gap hosts two non-degenerate eigenvalues. Panel \textbf{c},\textbf{d}: probability density plots of the states corresponding to the eigenvalues n${}^\circ 9$ and n${}^\circ 10$ in Panel \textbf{c} respectively (cf. red and orange dot). The first is localized at the domain wall, while the latter is localized at the left end of the strip, \textit{i.e.} at the termination of the topologically non-trivial side.}
    \label{fig:VarPot_N_6-M_0.5_1.3}
\end{figure}

\begin{figure}[p]
    \centering
    \includegraphics[width=\linewidth]{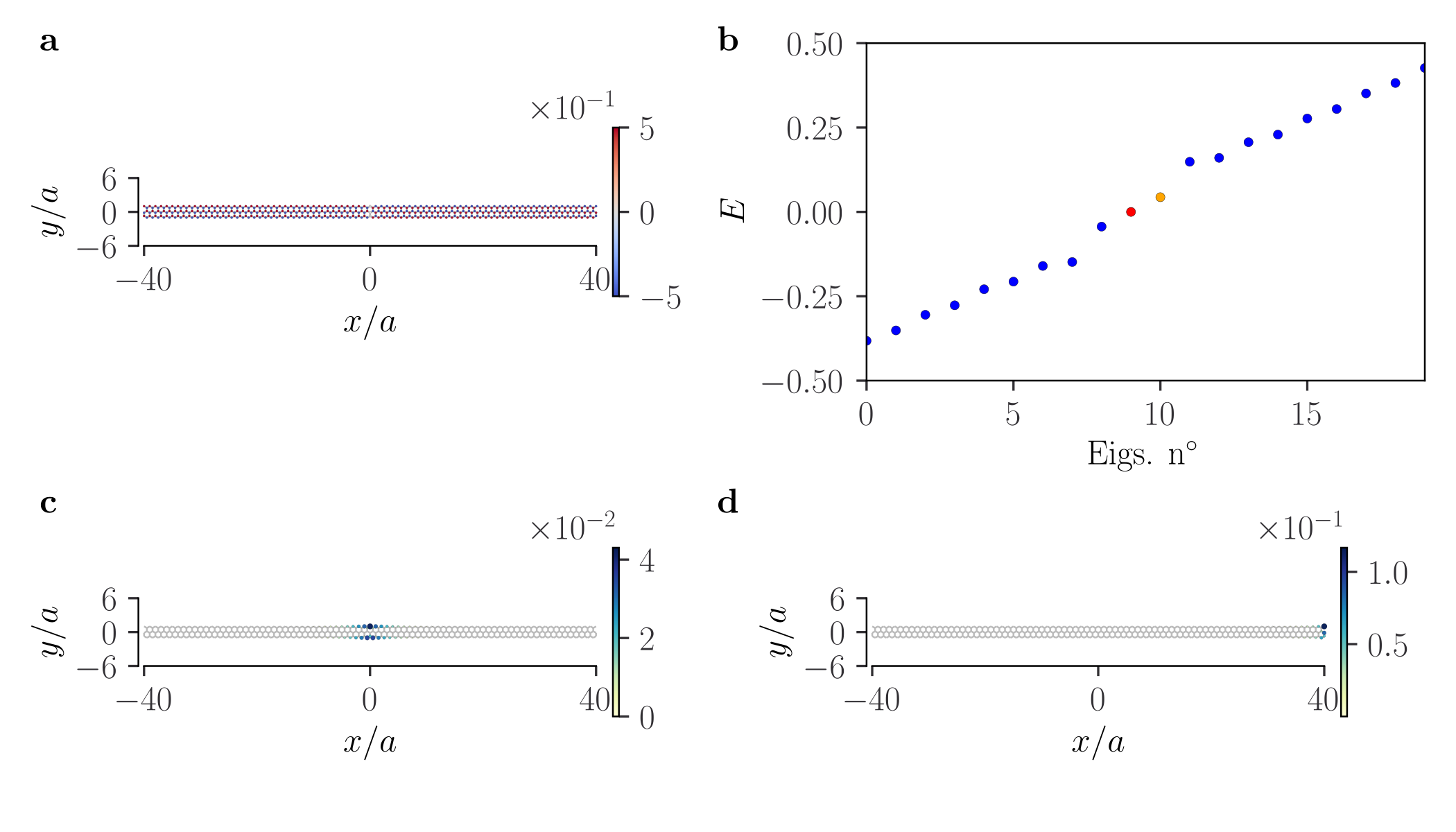}
    \caption{Results obtained through numerical diagonalization for the Haldane model on a zigzag nanoribbon (OBC) with $N_y=6$ and $L=80a$ with the staggered on-site potential interpolating between $-0.5$ and $0.5$. Panel \textbf{a}: representation of the on-site potential, changing from $m=-0.5$ for $x<0$ to $m=0.5$ for $x>0$. According to the phase diagram in the main text the two values of $m$ pertain to the two distinct contiguous regions, having the Zak phase differing of $\pi$. However both regions are topologically non trivial in terms of end states. Panel \textbf{b}: low energy spectrum of the model. The gap hosts three non-degenerate eigenvalues. Panel \textbf{c},\textbf{d}: probability density plots of the states corresponding to the eigenvalues n${}^\circ 9$ and n${}^\circ 10$ in Panel \textbf{c} respectively (cf. red and orange dot). The first is localized at the domain wall, while the latter is localized at the right end of the strip, \textit{i.e.} at the termination of the topologically non-trivial side with $m=0.5$.}
    \label{fig:VarPot_N_6-M_-0.5_0.5}
\end{figure}
\clearpage

\section*{Supplementary Tables}
\begin{table}[h]
    \centering
    \begin{tabular}{c|c|c|}
        $N_y$ & $\varphi^{m<0}$& $\varphi^{m>0}$ \\
        \hline
        $8\ell$ & $0$ & $0$ \\
        $8\ell+2$ & $0$ & $\pi$ \\
        $8\ell+4$ & $\pi $& $\pi$ \\
        $8\ell+6 $& $\pi$ & $0$ \\
        \hline
    \end{tabular}
    \caption{Zak Phase in the large $m$ limit.}
    \label{tab:tab1}
\end{table}
\clearpage

\section*{Supplementary Notes}
\subsection*{Supplementary Note 1}
Given the Hamiltonian with PBC along the $\mathbf{a}_1$ direction in Eq.~(4) of the main text, the corresponding Schr\"odinger equation
\begin{equation}
    \mathcal{H}(k) \Psi = \epsilon \Psi,
\end{equation}
can be recast in a more explicit way by denoting
\begin{equation}
    \Psi_{2n-1}(k)= \psi_n^A(k), \ \Psi_{2n}(k)= \psi_n^B(k), \qquad n \in \{1,\ldots,N_y/2\}.
\end{equation}
In this notation we can can synthetically write 
\begin{align*}
    \varepsilon\psi_n^A(k) &= \bigg\{g_0\psi_n^B(k)+t_1\psi_{n-1}^B(k)+(g(k,\phi)+m)\psi_n^A(k)+g(k/2,-\phi)\big[\psi_{n+1}^A(k)+\psi_{n-1}^A(k)\big]\bigg\},\\
    \varepsilon\psi_n^B(k) &= \bigg\{g_0\psi_n^A(k)+t_1\psi_{n+1}^A(k)+(g(k,-\phi)-m)\psi_n^B(k)+g(k/2,\phi)\big[\psi_{n+1}^B(k)+\psi_{n-1}^B(k)\big]\bigg\},
\end{align*}
where $g(k,\phi) = 2t_2\cos(k+\phi)$ and $g_0=2t_1\cos(k/2)$. The above equations are to be read assuming that $\psi_{0}^{A/B}(k)=0=\psi_{\Tilde{N}+1}^{A/B}(k), \ \Tilde{N}=N_y/2$, because of open boundary conditions in the $y$ direction. With $\phi = \pi/2$ we get: $g(k,\pm \pi/2) = \mp 2t_2\sin(k)$. Thus, after renormalizing every constant in units of $t_1$, we have
{\small
\begin{align*}
    \varepsilon\psi_n^A(k) &= \bigg\{2\cos(k/2)\psi_n^B(k)+\psi_{n-1}^B(k)+(-2t_2\sin(k)+m)\psi_n^A(k)+2t_2\sin(k/2)\big[\psi_{n+1}^A(k)+\psi_{n-1}^A(k)\big]\bigg\},\\
    \varepsilon\psi_n^B(k) &= \bigg\{2\cos(k/2)\psi_n^A(k)+\psi_{n+1}^A(k)+(2t_2\sin(k)-m)\psi_n^B(k)-2t_2\sin(k/2)\big[\psi_{n+1}^B(k)+\psi_{n-1}^B(k)\big]\bigg\}.
\end{align*}
}
What we want to prove here is that for $\Tilde{N}=2M+1$ (\textit{i.e.} $N_y=4M+2$), $M\in \mathbb N$, and $m=0$, the dispersion relation is always metallic. By setting $k=\pi$ we get:
\begin{align*}
    \varepsilon\psi_n^A(\pi) &= \bigg\{\psi_{n-1}^B(\pi)+2t_2\big[\psi_{n+1}^A(\pi)+\psi_{n-1}^A(\pi)\big]\bigg\},\\
    \varepsilon\psi_n^B(\pi) &= \bigg\{\psi_{n+1}^A(\pi)-2t_2\big[\psi_{n+1}^B(\pi)+\psi_{n-1}^B(\pi)\big]\bigg\}.
\end{align*}
We now start looking for a zero energy solution. Enforcing $\varepsilon=0$, the above equations reduce to
\begin{align*}
    0&= \psi_{n-1}^B+2t_2\big[\psi_{n+1}^A+\psi_{n-1}^A\big],\\
    0 &= \psi_{n+1}^A-2t_2\big[\psi_{n+1}^B+\psi_{n-1}^B\big],
\end{align*}
where, to simplify the notation, we have omitted the fact that the $\psi_n^{A/B}(k)$ are calculated for $k=\pi$.

Imposing the boundary condition $\psi_{0}^{A/B}(k)=0$ we get
\begin{align*}
    0 &= 2t_2\psi_{2}^A,\\
    0 &= \psi_{2}^A-2t_2\psi_{2}^B, \\
    0 &= \psi_{1}^B+2t_2\big[\psi_{3}^A+\psi_{1}^A\big],\\
    0 &= \psi_{3}^A-2t_2\big[\psi_{3}^B+\psi_{1}^B\big],
    \\
    &\vdots
\end{align*}
so that in the end $\psi_{2\ell}^{A/B}=0$ for any $\ell \in \mathbb N$. On the other hand, the boundary condition $\psi_{\Tilde{N}+1}^{A/B}(k)=0$ imposes no more conditions if $\Tilde{N}=2M+1$, since then $\Tilde{N}+1$ is even, while for $\Tilde{N}=2M$ it forces $\psi_{2\ell+1}^{A/B}=0$ for any $\ell \in \mathbb N$ as well. Thus we conclude that if $N_y=4M$ no zero energy solution exists for $m=0$. Note that the dispersion relation can only be gapless at $k=\pi$, since that is the position of the Dirac point for the edge states of the Haldane model on a strip with zigzag edges (see Fig.~1\textbf{b} of the main text).

For the $\Tilde{N}=2M+1$ case we look for a solution of the following kind:
\begin{align*}
    \psi_{2\ell}^{A/B} &=0, \\
    \psi_{2\ell+1}^{A/B} &= \xi^\ell \psi_{1}^{A/B}.
\end{align*}
The first condition is coherent with the boundary conditions, as already discussed. About the second one, by substituting $n=2\ell$ into the equations one gets
\begin{align*}
    0&= \psi_{2\ell-1}^B+2t_2\big[\psi_{2\ell+1}^A+\psi_{2\ell-1}^A\big]=\xi^{\ell-1}\psi_{1}^B+2t_2\big[\xi^{\ell}\psi_{1}^A+\xi^{\ell-1}\psi_{1}^A\big],\\
    0 &= \psi_{2\ell+1}^A-2t_2\big[\psi_{2\ell+1}^B+\psi_{2\ell-1}^B\big]=\xi^{\ell}\psi_{1}^A-2t_2\big[\xi^{\ell}\psi_{1}^B+\xi^{\ell-1}\psi_{1}^B\big],
\end{align*}
which upon dividing by $\xi^{\ell-1}$ yield
\begin{align*}
    0&=\psi_{1}^B+2t_2\big[\xi\psi_{1}^A+\psi_{1}^A\big],\\
    0 &= \xi\psi_{1}^A-2t_2\big[\xi\psi_{1}^B+\psi_{1}^B\big].
\end{align*}
Solving the system we get
\begin{equation}
    \psi_1^B=\dfrac{\xi}{2t_2(1+\xi)}\psi_1^A \implies \dfrac{\xi}{2t_2(1+\xi)}+2t_2(1+\xi)=0 \implies \xi + 4t_2^2(1+\xi)^2=0.
\end{equation}
Solving the equation for $\xi$ we get:
\begin{equation}
    \xi^2 + \xi (1/(4t_2^2) + 2) + 1 = 0 \implies \xi_{\pm} = - (1/(8t_2^2) + 1) \pm \sqrt{(1/(8t_2^2) + 1)^2-1}.
\end{equation}
Thus, we find two solutions with zero energy at $k=\pi$ for $m=0$ in the $N_y=4M+2$ case, as claimed in the main text. Finally, we observe that the product of the solutions is $1$, meaning that they have the same sign, and their sum equals $-(1/(4t_2^2) + 2)$,  meaning that they are both negative, one bigger and one smaller then unity in modulus respectively. These two degenerate solutions correspond to a state localized on the upper edge and to a state localized on the lower one, as expected.

\subsection*{Supplementary Note 2}
Let $N_y$ be the number of sites in the vertical direction (or, equivalently in the $\vec{a}_2$ direction).
In the large $m$ limit ($m\gg t_1,t_2$), the Hamiltonian can be brutally approximated as 
\begin{equation}
    \mathcal{H}(k)= \mathrm{diag}(m,-m,m,-m,\ldots).
\end{equation}
If the system is at half filling, the occupied states will be those localized on the sublattice with negative value of the on-site potential, \textit{i.e.} the $B$ sublattice for $m>0$ and the $A$ sublattice for $m<0$.

Let us start by considering $m>0$. For each $k \in \{0,\frac{2\pi}{Na}, \ldots,\frac{2\pi}{Na}(N-1)\}$ a basis of eigenstates for the $N_y/2\ (\equiv \Bar{n})$  occupied bands is given by
\begin{equation}
    \Psi^{m>0}_{1_\ell}(k) = \delta_{2,\ell},\ \ldots, \ \Psi^{m>0}_{\Bar{n}_\ell}(k) = \delta_{2\Bar{n},\ell}, \qquad \ell \in 1,\ldots,N_y,
\end{equation}
where the $\Psi_n(k)$ are $N_y$-dimensional vectors and the index $\ell$ refers to their components.
Thus, for $j=0,\ldots, N-2$, the $S$ matrix is trivially given by
\begin{equation}
    S^{m>0}(k_j,k_{j+1})= \mathbb I_{\Bar{n}},
\end{equation}
so that $\det[S^{m>0}(k_j,k_{j+1})]=+1 \ \forall j \in \{0,\ldots, N-2\}$.
However, from the periodic gauge enforcing,
\begin{equation}
    \Psi^{m>0}_{n_\ell}(k_N) = \mathrm{e}^{-\mathrm{i}\frac{2\pi}{a}t_\ell}  \Psi^{m>0}_{n_\ell}(k_0),
\end{equation}
where $t_\ell$ is the $x$-position of the $\ell$-th site in the strip unit cell. With our conventions (see main text) we get
\begin{align*}
    \mathrm{e}^{-\mathrm{i}\frac{2\pi}{a}t_{4r}} &= 1, \\
    \mathrm{e}^{-\mathrm{i}\frac{2\pi}{a}t_{4r+1}} &= 1, \\
    \mathrm{e}^{-\mathrm{i}\frac{2\pi}{a}t_{4r+2}} &= -1, \\
    \mathrm{e}^{-\mathrm{i}\frac{2\pi}{a}t_{4r+3}} &= -1.
\end{align*}
Thus, for $m>0$ we get
\begin{align*}
    \Psi^{m>0}_{2n_\ell}(k_N) &= \delta_{4n,\ell}, \\
    \Psi^{m>0}_{{2n-1}_\ell}(k_N) &= -\delta_{4n-2,\ell},
\end{align*}
or, equivalently
\begin{equation}
    \Psi^{m>0}_{n_\ell}(k_N) = (-1)^n \delta_{2n,\ell}.
\end{equation}
Recalling that at half filling $N_y=2\Bar{n}$, we have
\begin{equation}
    [S^{m>0}(k_{N-1},k_{N})]_{rs}= \Psi^{m>0\dagger}_r(k_{N-1})\Psi^{m>0}_s(k_N) = (-1)^s \sum_{\ell=1}^{N_y}\delta_{2r,\ell}\delta_{\ell, 2s}= (-1)^{r}\delta_{rs},
\end{equation}
and therefore the determinant turns out to depend on the value of $\Bar{n}$
\begin{equation}
    \det[S^{m>0}(k_{N-1},k_{N})]= 
    \begin{cases}
        +1 &\Bar{n}= 4\ell \\
        -1 &\Bar{n} = 4\ell +1 \\
        -1 &\Bar{n} = 4\ell +2 \\
        +1 &\Bar{n} = 4\ell + 3 \\
    \end{cases}
\end{equation}

We now turn to the $m<0$ case.
For each $k \in \{0,\frac{2\pi}{Na}, \ldots,\frac{2\pi}{Na}(N-1)\}$ a basis of eigenstates of the the $N_y/2\ (\equiv \Bar{n})$ occupied bands is now given by
\begin{equation}
    \Psi^{m<0}_{1_\ell}(k) = \delta_{1,\ell},\ \ldots, \ \Psi^{m<0}_{\Bar{n}_\ell}(k) = \delta_{2\Bar{n}-1,\ell}, \qquad \ell \in 1,\ldots,N_y.
\end{equation}
Again for $j=0,\ldots, N-2$, the $S$ matrix is trivially given by
\begin{equation}
    S^{m<0}(k_j,k_{j+1})= \mathbb I_{\Bar{n}},
\end{equation}
so that $\det[S^{m<0}(k_j,k_{j+1})]=+1 \ \forall j \in \{0,\ldots, N-2\}$. However, enforcing the periodic gauge for $m<0$ yields ($\Psi^{m<0}_{n_\ell}(k)=\delta_{2n-1,\ell}$)
\begin{align*}
    \Psi^{m<0}_{2n_\ell}(k_N) &= -\delta_{4n-1,\ell}, \\
    \Psi^{m<0}_{{2n-1}_\ell}(k_N) &= +\delta_{4n-3,\ell},
\end{align*}
or, equivalently
\begin{equation}
    \Psi^{m<0}_{n_\ell}(k_N) = (-1)^{n-1} \delta_{2n-1,\ell}.
\end{equation}
Then, proceeding as in the $m>0$ case
\begin{equation}
    [S^{m<0}(k_{N-1},k_{N})]_{rs} = (-1)^{r-1}\delta_{rs} \implies \det[S(k_{N-1},k_{N})]= 
    \begin{cases}
        +1 &\Bar{n}= 4\ell \\
        +1 &\Bar{n} = 4\ell +1 \\
        -1 &\Bar{n} = 4\ell +2 \\
        -1 &\Bar{n} = 4\ell + 3 \\
    \end{cases}
\end{equation}

We have found that in the large $m$ limit the sign of  $\prod_{j=0}^{N-1} \det[S(k_{j},k_{j+1})] $ is entirely determined by the sign of $\det[S(k_{N-1},k_{N})]$. Knowing that, we can easily determine the Zak phase, which is given by
\[
    \varphi = -\Im \log \prod_{j=0}^{N-1} \det[S(k_{j},k_{j+1})], \qquad \varphi \in (-\pi,\pi].
\]
The resulting values of the Zak phase in the large (positive and negative) $m$ limit are reported as a function of $N_y$ ($N_{y}=2\Bar{n}$) in Supplementary Table~\ref{tab:tab1}. In particular, by inspecting Supplementary Table~\ref{tab:tab1} we conclude that for $N_y=4M+2$ the Zak phase differs of $\pi$ between the $m\to \pm \infty$ limits, while this is not the case for $N_y=4M$.

\subsection*{Supplementary Note 3}
In order to test the robustness of the bound states, we assess their stability against the introduction of random on-site disorder of the following form
\begin{equation}
    \mathcal{H}_\text{RN} = \sum_{i} \omega_i c_i^\dagger c_i,
\end{equation}
where the sum runs over all lattice sites, the operator $c_i$ destroys a fermion at site $i$ and $\omega_i$ is a real number randomly extracted from the interval $[-V,V]$, $V>0$. In Supplementary Figure~\ref{fig:noise_N_2} we compare two Haldane zigzag nanoribbons (OBC geometry) with $N_y=4$ and $L=20a$ without (Panels \textbf{a}, \textbf{c}, \textbf{e}, \textbf{g}) and with (Panels \textbf{b}, \textbf{d}, \textbf{f}, \textbf{h}) disorder respectively; an analogous comparison for a strip with $N_y=6$ and $L=40a$ is reported in Supplementary Figure~\ref{fig:noise_N_3}. The parameters of the Haldane model are set as in the main text to $t_1=1, \ t_2=0.3, \ \phi=\frac{\pi}{2}$, while the maximum intensity of the on-site disorder is chosen as $V=0.2$. The mass term is set to $m=0$ for the $N_y=4$ strip and to $m=0.5$ for the $N_y=6$ one, so that both systems are in their topological phase and host end states.

For both configurations, we find that the effect of the random on-site disorder is to break the degeneracy between the end states, which, however, survive to the perturbation.
In Panels \textbf{a} and \textbf{b} of Supplementary Figure~\ref{fig:noise_N_2} (Supplementary Figure~\ref{fig:noise_N_3}) we report a graphical representation of the on-site potential in absence and presence of the random noise term for the strip with $N_y=4$ ($N_y=6$). In Panels \textbf{c} and \textbf{d} of Supplementary Figure~\ref{fig:noise_N_2} (Supplementary Figure~\ref{fig:noise_N_3}) we compare the corresponding low energy spectra: even in the presence of noise, two eigenmodes (marked in red and orange) are present inside the energy gap. However, in contrast to the unperturbed case, the in-gap eigenvalues in Supplementary Figure~\ref{fig:noise_N_2}\textbf{d} (Supplementary Figure~\ref{fig:noise_N_3}\textbf{d}) are slightly split in energy. In Panels \textbf{e} and \textbf{g} of Supplementary Figure~\ref{fig:noise_N_2} (Supplementary Figure~\ref{fig:noise_N_3}) are reported the probability density plots of the eigenstates corresponding to the eigenvalues marked in red and orange in Panel \textbf{c}, while in Panels \textbf{f} and \textbf{h} of Supplementary Figure~\ref{fig:noise_N_2} (Supplementary Figure~\ref{fig:noise_N_3}) are reported the ones corresponding to the eigenvalues marked in red and orange in Panel \textbf{d}. By comparing them one can see that, as a consequence of the energy splitting, the two bound states, which in absence of noise are degenerate in energy and equally localized on both ends, localize themselves on one end of the strip each.

Similar results to those discussed here hold for strips of greater width or different random noise configurations. However, being that the splitting between the end states is due to the value of the on-site disorder at the two ends of the strip (\textit{i.e.} where the bound states are localized), the splitting of the eigenmodes (and so their localization pattern) may differ for different noise configurations.

\subsection*{Supplementary Note 4}
To derive an effective analytical expression for the coupling term $\mathcal{M}$ in Eq.~9 of the main text in the case $m=0$, we start by considering the coupling between two one dimensional chains, formed by the edge sites of the zigzag nanoribbon. We introduce an exponentially suppressed coupling between the two chains, of the form
\[
    t_{ij} = \mathrm{e}^{-d_{ij}/\xi},
\]
with $d_{ij}$ the distance between the site $i$ of one edge and the site $j$ of the other and $\xi$ some length scale.

Also, we set a sharp ``first-neighbor'' cutoff, effectively limiting the interaction to the edge sites that in the full strip are connected by the minimum amount of first neighbor hoppings (see Supplementary Figure~\ref{fig:eff_coupl_scheme}). According to this criterion, the number of coordination is easily found to be $n_c = \frac{N_y}{2}+1$. Thus, the interaction Hamiltonian is written as
\begin{align*}
    &N_y=4M: && H_{\text{int}} = \Tilde{\Delta}\sum_{\ell=1}^{N}\sum_{j=-(n_c-1)/2}^{(n_c-1)/2} \mathrm{e}^{-(\sqrt{w^2+(ja)^2}-w)/\xi} a^\dagger_\ell b_{\ell+j} + \text{h.c.}, \\
    &N_y=4M+2: && H_{\text{int}} = \Tilde{\Delta}\sum_{\ell =1}^{N}\sum_{j=-n_c/2}^{n_c/2-1} \mathrm{e}^{-(\sqrt{w^2+(ja+a/2)^2}-w)/\xi} a^\dagger_\ell b_{\ell+j} +\text{h.c.},
\end{align*}
where $\Tilde{\Delta}$ is some arbitrary energy scale and $w$ the strip width. The operator $a_\ell$ ($b_\ell$) destroys a fermion on the site $\ell $ of the lower (upper) edge, corresponding to a $A$ ($B$) site.
The strip width $w$ can be expressed in terms of $N_y$ as (first neighbor distance set to unity)
\[
    w = \dfrac{\sqrt{3}}{2}\dfrac{N_y}{2}-\dfrac{1}{\sqrt{3}}.
\]
About the length scale $\xi$, it is reasonable to assume that it should be of the same order of magnitude of the localization length of the edge states. This is given analytically for $k=\pi$ (assuming $m=0$ and $\phi=\frac{\pi}{2}$) in~\cite{Doh_2013, Cano_2013}
\[
    \xi_{\text{loc}}(\pi) = \dfrac{\sqrt{3}}{2}\left[\log\left\{\sqrt{1+\left(\frac{t_1}{4t_2}\right)^2} +\frac{t_1}{4t_2}\right\}\right]^{-1},
\]
and is shown to be more or less constant around the Dirac point and for small values of $m$. More specifically, since the coupling results from the interaction between the two opposite chiral modes, each of which has a decay length $\sim\xi_{\text{loc}}(\pi)$, we pick $\xi=2\xi_{\text{loc}}(\pi)$ for the plots in Figs.~5\textbf{a} to 5\textbf{d} of the main text.

By Fourier transforming we find the interacting Hamiltonian in $k$-space. First we rewrite expressing $n_c$ in terms of $M$:
\begin{align*}
    &N_y=4M: && H_{\text{int}} = \Tilde{\Delta}\sum_{\ell =1}^{N}\sum_{j=-M}^{M} \mathrm{e}^{-(\sqrt{w^2+(ja)^2}-w)/\xi} a^\dagger_\ell b_{\ell +j} + \text{h.c.}, \\
    &N_y=4M+2: && H_{\text{int}} = \Tilde{\Delta}\sum_{\ell =1}^{N}\sum_{j=-(M+1)}^{M} \mathrm{e}^{-(\sqrt{w^2+(ja+a/2)^2}-w)/\xi} a^\dagger_\ell b_{\ell+j} +\text{h.c.}
\end{align*}
The Fourier transformation is defined as
\begin{align}
    a_{\ell}&= \dfrac{1}{\sqrt{N}}\sum_{k} \mathrm{e}^{-\mathrm{i}k\ell a}a(k), \\
    b_{\ell}&= \dfrac{1}{\sqrt{N}}\sum_{k} \mathrm{e}^{-\mathrm{i}k(\ell a+\delta)}b(k) ,
\end{align}
with $\delta = 0$ if $N_y=4M$ and $\delta = 1/2$ if $N_y=4M+2$. Thus
\begin{align*}
    &N_y=4M: && H_{\text{int}} = \frac{\Tilde{\Delta}}{N}\sum_{\ell=1}^{N} \sum_{k,q}\sum_{j=-M}^{M} \mathrm{e}^{-(\sqrt{w^2+j^2}-w)/\xi} \mathrm{e}^{\mathrm{i}k\ell}\mathrm{e}^{-\mathrm{i}q(\ell+j)}a^\dagger(k)b(q) + \text{h.c.} \\
    &N_y=4M+2: && H_{\text{int}} = \frac{\Tilde{\Delta}}{N}\sum_{\ell=1}^{N}\sum_{k,q}\sum_{j=-(M+1)}^{M} \mathrm{e}^{-(\sqrt{w^2+(j+1/2)^2}-w)/\xi} \mathrm{e}^{\mathrm{i}k\ell}\mathrm{e}^{-\mathrm{i}q(\ell+j+1/2)} a^\dagger(k)b(q)+\text{h.c.}
\end{align*}
Summing over $\ell$ and $q$:
\begin{align*}
    &N_y=4M: && H_{\text{int}} = \Tilde{\Delta}\sum_{k}\sum_{j=-M}^{M} \mathrm{e}^{-(\sqrt{w^2+j^2}-w)/\xi} \mathrm{e}^{-\mathrm{i}kj}a^\dagger(k)b(k) + \text{h.c.} \\
    &N_y=4M+2: && H_{\text{int}} = \Tilde{\Delta}\sum_{k}\sum_{j=-(M+1)}^{M} \mathrm{e}^{-(\sqrt{w^2+(j+1/2)^2}-w)/\xi} \mathrm{e}^{-\mathrm{i}k(j+1/2)} a^\dagger(k)b(k) +\text{h.c.}
\end{align*}
Now we consider the $N_y=4M$ case first. We have
\[
    H_{\text{int}} = \Tilde{\Delta}\sum_{k} [1 + \sum_{j=1}^{M} 2\cos(kj)\mathrm{e}^{-(\sqrt{w^2+j^2}-w)/\xi}]a^\dagger(k)b(k) + \text{h.c.},
\]
yielding
\begin{equation}
    \mathcal{H}^{\text{edge}}_{\text{int}}(k)= \Tilde{\Delta} [1 + \sum_{j=1}^{M} 2\cos(kj)\mathrm{e}^{-(\sqrt{w^2+j^2}-w)/\xi}] \tau_x.
\end{equation}
Proceeding analogously for the $N_y=4M+2$ case
\[
    \begin{split}
        H_{\text{int}} &=\Tilde{\Delta}\sum_{k}\left\{\sum_{j=0}^{M} \mathrm{e}^{-(\sqrt{w^2+(j+1/2)^2}-w)/\xi} \mathrm{e}^{-\mathrm{i}k(j+1/2)} a^\dagger(k)b(k) +\text{h.c.} \right.\\
        &\left.+ \sum_{j=1}^{M+1} \mathrm{e}^{-(\sqrt{w^2+(-j+1/2)^2}-w)/\xi} \mathrm{e}^{-\mathrm{i}k(-j+1/2)} a^\dagger(k)b(k) +\text{h.c.}\right\}\\
        &= \Tilde{\Delta}\sum_{k}\left\{\sum_{j=0}^{M} \mathrm{e}^{-(\sqrt{w^2+(j+1/2)^2}-w)/\xi} \mathrm{e}^{-\mathrm{i}k(j+1/2)} a^\dagger(k)b(k) +\text{h.c.} \right.\\
        &\left.+ \sum_{j=0}^{M} \mathrm{e}^{-(\sqrt{w^2+(j+1/2)^2}-w)/\xi} \mathrm{e}^{\mathrm{i}k(j+1/2)} a^\dagger(k)b(k) +\text{h.c.}\right\}\\
        &= \Tilde{\Delta}\sum_{k}\sum_{j=0}^{M} \mathrm{e}^{-(\sqrt{w^2+(j+1/2)^2}-w)/\xi} 2\cos(k(j+1/2)) a^\dagger(k)b(k) +\text{h.c.} \\
    \end{split}
\]
This yields
\begin{equation}
    \mathcal{H}^{\text{edge}}_{\text{int}}(k)= \Tilde{\Delta} \sum_{j=0}^{M} 2\cos(k(j+1/2)) \mathrm{e}^{-(\sqrt{w^2+(j+1/2)^2}-w)/\xi}  \tau_x.
\end{equation}
Thus, if we identify $\mathcal{H}^{\text{edge}}_{\text{int}}(k)= \mathcal{M}^{\text{teo}}(k)\tau_x$ in the end we have
\begin{align}
    &N_y=4M: && \mathcal{M}^{\text{teo}}(k) = \Tilde{\Delta} [1 + \sum_{j=1}^{M} 2\cos(kj)\mathrm{e}^{-(\sqrt{w^2+j^2}-w)/\xi}],
    \label{eq:delta_even}
    \\
    &N_y=4M+2: && \mathcal{M}^{\text{teo}}(k) = \Tilde{\Delta} \sum_{j=0}^{M} 2\cos(k(j+1/2)) \mathrm{e}^{-(\sqrt{w^2+(j+1/2)^2}-w)/\xi}.
    \label{eq:delta_odd}
\end{align}

Moreover, as explained in the main text, we can numerically retrieve the mass term $\mathcal{M}$ as a function of the model parameters for a given $N_y$ via
\begin{equation}
    |\mathcal{M}(m;k)|= \sqrt{E_{N_y}(m; k)^2-E_{\infty}(m; k)^2},
    \label{eq:mass_num}
\end{equation}
where $E_{N_y}(m; k)$ is the-- exact, numerically computed --lowest positive energy band for a strip of width $N_y$ (all the other parameters of the model fixed), and $E_{\infty}(m; k)$ is the lowest positive energy band for a model with the same identical parameters but, on a much wider strip ($N_y\to \infty)$.

In Supplementary Figures~\ref{fig:Mk_t2_015-num}, \ref{fig:Mk_t2_02-num} and \ref{fig:Mk_t2_03-num} are shown the plots of $\mathcal{M}$ as a function of $k$ and for different values of the staggered mass $m$-- each column corresponding to one of the widths considered in the main text ($N_y=4,6,8,10$) --for $t_2=0.15$, $t_2=0.2$ and $t_2=0.3$ respectively. The reference spectrum ($E_{\infty}(m; k)$ in Eq.~\ref{eq:mass_num}) was obtained considering a strip with $N_y =202$ sites in the vertical direction.

The general features that we can retrieve from these plots are the following: The number of nodes in the effective mass increases of exactly one from one zigzag strip to the next larger one; for $m=0$ the effective mass has a node at $\pi$ when $N_y=4M+2, \ M \in \mathbb{Z}$; as $m$ is increased (to positive values) the profile of $\mathcal{M}$ moves to the right in the Brillouin zone; the higher the value of $t_2$ the more the profile of $\mathcal{M}$ spreads in $k$-space. The points in $k$ space where $\mathcal{M}$ stops oscillating and steeply goes up correspond to the points where the edge states of the wide (reference) strip merge with the bulk states. That is, where the low energy theory ceases to work.

\subsection*{Supplementary Note 5}
Here we report some more phenomenology regarding the Jackiw-Rebbi like bound states occurring at the interface between sides of the strip belonging to topologically distinct regions of the phase diagram. We set the parameters as in the main text to $t_1=1, \ t_2=0.3, \ \phi=\frac{\pi}{2}$.
In the main text we considered a scenario with $N_y=6$ and with $m$ interpolating between two regions of the phase diagram both trivial with respect to the presence of end states, but with the Zak phase differing of $\pi$. Here we consider a more detailed, though necessarily not exhaustive, phenomenology.

In Supplementary Figure~\ref{fig:VarPot_N_4-M_-0.3_0.3} we consider a zigzag Haldane nanoribbon with $N_y=4$ and $L=40a$, with the staggered on-site potential $m$ interpolating between $-0.3$ and $0.3$ (Panel \textbf{a}). With reference to the phase diagram in Fig.~2\textbf{a} of the main text, the two sides of the strip belong to the same region of the phase diagram: Therefore, no bound states at the domain wall would be expected. Indeed, from numerical diagonalization we find that two isolated degenerate eigenvalues occur inside the gap (Panel \textbf{b}). These correspond to two bound states which, according to the probability density plots in Panels \textbf{c}, \textbf{d}, are equally localized on the two ends of the strip. These are actually expected since both sides of the strip are in the topologically non trivial region (cf. Fig.~2\textbf{a} of the main text).

In Supplementary Figure~\ref{fig:VarPot_N_6-M_0.5_1.3} we consider a zigzag Haldane nanoribbon with $N_y=6$ and $L=80a$, with the staggered on-site potential $m$ interpolating between $0.5$ and $1.3$ (Panel \textbf{a}). With reference to the phase diagram in Fig.~2\textbf{b} of the main text, the two sides of the strip belong to distinct contiguous regions of the phase diagram: in terms of end states $m=0.5$ belongs to the topologically non-trivial region, while $m=1.3$ to the trivial (and connected to the atomic limit) one. Since the Zak phase differs of $\pi$ between the two regions, a bound state is expected at the domain wall. The low energy spectrum obtained via numerical diagonalization is reported in Panel \textbf{b}. We find that two isolated eigenvalues occur inside the gap: The lower one in energy (coloured in red) corresponds to an eigenstate localized at the domain wall in the on-site potential, as shown by the probability density plot in Panel \textbf{c}. The one at higher energy instead (coloured in orange), is localized at the left end of the strip (Panel \textbf{d}): this is expected since the left side, having $m=0.5$, pertain to a topological region (cf. Fig.~2\textbf{b} of the main text).

In Supplementary Figure~\ref{fig:VarPot_N_6-M_-0.5_0.5} we consider a zigzag Haldane nanoribbon with $N_y=6$ and $L=80a$, with the staggered on-site potential $m$ interpolating between $-0.5$ and $0.5$ (Panel \textbf{a}). With reference to the phase diagram in Fig.~2\textbf{b} of the main text, the two sides of the strip belong to distinct contiguous regions of the phase diagram: in terms of end states, both sides belong to topologically non trivial regions of the phase diagrams. Despite this, since the Zak phase differs of $\pi$ between the two regions, a bound state is expected at the domain wall (similarly to the case considered in the main text). The low energy spectrum obtained via numerical diagonalization is reported in Panel \textbf{b}. We find that three isolated eigenvalues occur inside the gap: The mid one in energy (coloured in red) corresponds to an eigenstate localized at the domain wall in the on-site potential, as shown by the probability density plot in Panel \textbf{c}. The one at higher energy instead (coloured in orange), is localized at the right end of the strip (Panel \textbf{d}). Finally, the lower one in energy correspond to an eigenstate localized at the left end of the strip (the corresponding probability density plot is not reported for brevity). Again, these latter two bound states are expected, since the right (left) side, having $m=0.5$ ($m=-0.5$), pertain to a topological region (cf. Fig.~2\textbf{b} of the main text).